\documentclass[12pt]{article}
\usepackage{caption}
\usepackage{amssymb}
\usepackage{amsmath}
\usepackage{hyperref}
\usepackage{graphicx}
\usepackage{float}
\usepackage{bm}
\usepackage{latexsym}
\usepackage{subfig}
\usepackage{mathtools}
\usepackage{booktabs}
\usepackage{tabularx}
\title{\bf Thermodynamic Topology of 4D Charged AdS Black Holes with $F^{\alpha\beta}F^{\gamma\lambda}R_{\alpha\gamma\beta\lambda}$ Coupling}
\author{Mehdi Sadeghi\thanks{Email: mehdi.sadeghi@abru.ac.ir}\,\,\, and \,\,Faramaz Rahmani\thanks{Corresponding author: faramarz.rahmani@abru.ac.ir}\hspace{2mm}\\
{\small {\em Department of Physics, Faculty of Basic Sciences,}}\\
{\small {\em Ayatollah Boroujerdi University, Boroujerd, Iran}}
}
\date{\today}

\begin{document}

\maketitle

\begin{abstract}
We investigate the thermodynamic phase transitions of a four-dimensional charged anti-de Sitter black hole endowed with a non-minimal coupling of the form $F^{\alpha\beta}F^{\gamma\lambda}R_{\alpha\gamma\beta\lambda}$. Using perturbative methods, we derive a consistent black hole solution and analyze its thermodynamics through both conventional equilibrium techniques and a topological defect classification approach. The system displays van der Waals-like critical behavior, with a swallow-tail structure in the free energy and distinct phase branches. The topological analysis independently confirms the existence of critical points and classifies the system within the universal topological scheme for black hole thermodynamics.
\end{abstract}



%--------------------------------------------------------------------------
\section{Introduction} \label{intro}

Einstein's theory of general relativity revolutionized our understanding of gravity by interpreting it as the curvature of spacetime caused by matter and energy \cite{Einstein1915}. 
In this geometric description, black holes arise as vacuum solutions to Einstein's equations \cite{Schwarzschild1916}. 
These objects serve as theoretical laboratories for exploring the interplay between gravity, thermodynamics, and quantum mechanics. 
The discovery of black hole thermodynamics with entropy proportional to horizon area and temperature proportional to surface gravity, was pivotal \cite{Bekenstein1973,Hawking1975}. 
These results motivated the holographic principle \cite{tHooft1993,Susskind1995} and microscopic explanations in quantum gravity frameworks such as string theory and loop quantum gravity \cite{Maldacena1998,Ashtekar2000}.

Among black hole solutions, those embedded in AdS spacetime have garnered particular interest
due to their thermodynamic stability and rich phase structure \cite{HawkingPage1983,Witten1998}.
The AdS/CFT correspondence provides a dual interpretation of the Hawking--Page transition
as a confinement--deconfinement phase transition in gauge theories \cite{Maldacena1998,Witten1998}.
AdS black holes have since become central tools for studying strongly coupled quantum systems,
hydrodynamic transport, and emergent spacetime geometry \cite{Policastro2001,Hubeny2011}.

The thermodynamic landscape of AdS black holes has been significantly enriched by considering various nonlinear extensions to standard Einstein-Maxwell theory. These nonlinear models introduce novel critical phenomena and phase structures inaccessible in the linear regime. Among the most influential AdS nonlinear models are:
\begin{itemize}
    \item \textbf{Born--Infeld electrodynamics}, originally proposed to eliminate point-charge divergences,
    introduces maximal field strength effects and gives rise to Van der Waals--like liquid--gas phase
    transitions, reentrant phase transitions, and triple points in charged AdS black holes
    \cite{BornInfeld1934,BIAdS2012,TriplePoint2014}.
    
    \item \textbf{Gauss--Bonnet and Lovelock gravities}, representing higher-curvature corrections motivated
    by string theory, yield non-mean-field critical exponents and multiple horizon branches while preserving
    second-order field equations \cite{Lovelock1971,GBAdS2012}.
    
    \item \textbf{Massive gravity theories}, particularly the ghost-free dRGT model, allow the graviton
    mass to act as an additional thermodynamic variable, enabling quadruple points and novel phase
    boundaries for AdS black holes \cite{dRGT2011,MassiveBH2015}.
    
    \item \textbf{Horndeski and generalized scalar-tensor theories}, comprising the most general
    second-order scalar--tensor interactions, produce nontrivial scalar hair and modified critical
    behavior in AdS black hole thermodynamics \cite{Horndeski1974,HorndeskiBH2016}.
    
    \item \textbf{Non-minimal curvature--gauge coupling models}, including interactions of the form
    $F^{\alpha\beta}F^{\gamma\lambda}R_{\alpha\gamma\beta\lambda}$ considered in this work, directly
    couple electromagnetic fields to spacetime geometry and are motivated by quantum gravity,
    effective field theory, and curvature singularity regularization
    \cite{DrummondHathrell1980,NonMinimalBH2019}.
\end{itemize}

Each nonlinear extension modifies the thermodynamic phase space through distinct mechanisms whether through field strength saturation, higher-curvature terms, graviton mass effects, scalar interactions, or direct gauge-curvature couplings, yet all contribute to our understanding of critical phenomena in gravitational systems and their holographic duals.

The study of black hole thermodynamics has traditionally employed standard thermodynamic
analysis in the extended phase space, examining equations of state, free energies, heat
capacities, and stability criteria to identify phase transitions and critical phenomena
\cite{KubiznakMann2012,Altamirano2014}.
While this conventional approach has been remarkably successful in revealing Van der
Waals--like behavior, reentrant phase transitions, and triple points in various black hole
systems \cite{Altamirano2014,WeiLiu2014}, it often requires case-by-case analysis of specific
models and can obscure deeper structural patterns governing thermodynamic behavior across
different gravitational theories.

In recent years, a novel \textit{topological} perspective has emerged that offers a more
fundamental, model-independent classification of black hole phase transitions by treating
thermodynamic spaces as differentiable manifolds and examining their global topological
properties \cite{WeiLiu2020, Wei2021, Mo2022,Wei:2022dzw, Wei:2022dzw2,Yerra2022,Liu:2022aqt}.
The topological approach to black hole thermodynamics originated from applications of
Duan's topological current theory to defect classification in condensed matter systems,
and was later adapted to gravitational systems \cite{Duan1979,Duan2009}.
The central insight is that thermodynamic phase transitions correspond to topological
changes in the parameter space of solutions, with different phases representing distinct
topological sectors characterized by integer-valued topological charges \cite{WeiLiu2020}.
This methodology analyzes the zeros of thermodynamic functions as topological defects in
temperature--entropy or pressure--volume planes.
By constructing a thermodynamic potential whose extrema correspond to black hole solutions
and analyzing winding numbers around these extrema, one can assign topological charges to
different solution branches: stable black holes typically carry positive charges, unstable
ones negative charges, and phase transitions occur where topological charges change
\cite{WeiLiu2020,Mo2022}. Characteristically, topological investigations proceed by mapping the black hole parameter
space to a vector field whose zeros correspond to thermodynamic equilibrium states.
Phase transitions manifest as creations or annihilations of zero pairs with opposite
topological charges, analogous to vortex-antivortex pair production in condensed matter
systems \cite{WeiLiu2020,Ahmed:2022kyv}.

Compared to conventional thermodynamic analysis, the topological approach offers several
distinct advantages.
Traditional methods typically involve calculating free energy landscapes, examining
specific heat divergences, or analyzing $P$--$V$ criticality
\cite{KubiznakMann2012,Altamirano2014}.
The topological method, by contrast, focuses on global properties of the solution space,
often requiring only knowledge of the equations determining equilibrium states rather than
full analytic solutions \cite{Wei2021}.
Where conventional analysis might identify a phase transition through the merging of two
minima in free energy, topological analysis identifies the same phenomenon through the
cancellation of topological charges, providing a more fundamental geometric interpretation.
Furthermore, while standard thermodynamic stability analysis examines local perturbations, topological charges offer a global characterization of stability that can reveal subtle
features missed by local methods, such as metastable regions separated by topological barriers \cite{Mo2022}.

While the topological approach provides powerful new insights, it is most effective when used in conjunction with traditional thermodynamic methods. Topological classification offers a structural framework for understanding phase
transitions but generally does not provide quantitative predictions of critical temperatures or pressures, which still require conventional thermodynamic calculations.
Nevertheless, as a complement to traditional methods, topological analysis represents a significant advancement in our understanding of black hole thermodynamics.

In this work, we apply topological methods to investigate the phase structure of non-minimally coupled AdS black holes with $F^{\alpha\beta}F^{\gamma\lambda}R_{\alpha\gamma\beta\lambda}$ interaction. Our approach will compute topological charges for different solution branches, identify topological transitions corresponding to phase changes, and compare these results with conventional thermodynamic analysis of free energy and specific heat. 

%--------------------------------------------------------------------------
\section{$F^{\alpha\beta}F^{\gamma\lambda}R_{\alpha\gamma\beta\lambda}$-Coupled AdS Black Hole and Perturbative Solutions}\label{sec2}

We consider a non-minimal Einstein--Maxwell theory with a negative cosmological constant, described by the action~\cite{BalakinLemos2005,DereliSert2011,Mahapatra:2016dae,Sadeghi2023},

\begin{eqnarray}\label{action}
S=\int d^{4}  x\sqrt{-g} \bigg[\frac{1}{\kappa }(R-2\Lambda )-\frac{\alpha}{4}F_{\mu \alpha }F^{\mu \alpha} +\epsilon F^{\alpha\beta} F^{\gamma\lambda} R_{\alpha\gamma\beta\lambda} \bigg],
\end{eqnarray}
where $\kappa$ is the gravitational constant, $R$ is the Ricci scalar, $\Lambda =-\frac{3}{\ell^2}$ is the cosmological constant with $\ell$ the AdS radius, $\alpha$ is a dimensionless constant, and $F_{\mu\nu}$ is the Maxwell field strength tensor. The parameter $\epsilon$ is a dimensionful coupling that controls the non-minimal interaction between the gauge field and curvature.

The specific coupling term \(F^{\alpha\beta}F^{\gamma\lambda}R_{\alpha\gamma\beta\lambda}\) is motivated by several considerations in gravitational physics. From an effective field theory perspective, such non-minimal interactions naturally arise as quantum corrections when integrating out massive degrees of freedom in curved spacetime, representing the leading-order gauge–curvature coupling that preserves gauge invariance and general covariance \cite{Drummond:1980qv,Buchbinder:1992rb}. Geometrically, this term corresponds to a direct interaction between the electromagnetic field strength tensor and the full Riemann tensor, capturing more complete curvature–gauge interplay than simpler couplings to Ricci scalar or tensor alone. This makes it particularly relevant for studying phenomena where both electromagnetic and gravitational fields are strong and their mutual backreaction is significant, such as in the vicinity of charged black holes or in early universe cosmology.

Physically, the \(F^{\alpha\beta}F^{\gamma\lambda}R_{\alpha\gamma\beta\lambda}\) coupling modifies the propagation of electromagnetic waves in curved spacetime, potentially leading to birefringence effects and dispersion relations that depend on spacetime curvature, phenomena of interest in tests of fundamental physics using astrophysical observations \cite{Drummond:1980qv,Shore:2003zc}. Theoretically, such terms appear naturally in certain string theory compactifications and in the context of gravitational analogues of the Euler–Heisenberg Lagrangian, where they represent curvature-induced corrections to Maxwell electrodynamics \cite{Buchbinder:1992rb}. For black hole physics specifically, this coupling can regularize curvature singularities, modify thermodynamic properties in nontrivial ways, and introduce novel phase transitions that are inaccessible in minimally coupled Einstein–Maxwell theory \cite{Balakin:2015gpq,Sert:2020vmq}.

The perturbative treatment with small \(\epsilon\) allows us to systematically investigate these effects while maintaining analytic control over the solutions. For a consistent perturbative expansion, the dimensionful coupling constant 
$\epsilon$ ($[\epsilon] = L^2$) must be small relative to the natural curvature 
scale of the background, which is $\ell^{-2}$. This defines the dimensionless 
expansion parameter $\xi \equiv \epsilon/\ell^2$. Our analysis is conducted in 
the regime $|\xi| \ll 1$, where the non-minimal interaction 
$F^{\alpha\beta}F^{\gamma\lambda}R_{\alpha\gamma\beta\lambda}$ constitutes a 
controlled deformation of the standard Einstein--Maxwell thermodynamics.

The Maxwell tensor is defined as
\begin{align} \label{YM}
F_{\mu\nu} = \partial_\mu A_\nu - \partial_\nu A_\mu ,
\end{align}
where $A_\mu$ denotes the electromagnetic gauge potential. Varying the action~\eqref{action} with respect to the metric $g_{\mu\nu}$ yields the gravitational field equations
\begin{equation}\label{EOM1}
R_{\mu\nu} - \tfrac{1}{2} g_{\mu\nu} R + \Lambda g_{\mu\nu} = \kappa T^{\text{(eff)}}_{\mu\nu},
\end{equation}
with an effective energy--momentum tensor of the form
\begin{equation}
T^{\text{(eff)}}_{\mu\nu} = \alpha T^{\text{(M)}}_{\mu\nu} + \epsilon T^{(I)}_{\mu\nu}.
\end{equation}
Here
\begin{equation}
T^{\text{(M)}}_{\mu\nu} = \frac{1}{2} F_{\mu}^{\;\; \alpha} F_{\nu\alpha} - \tfrac{1}{8} g_{\mu\nu} F_{\alpha\beta} F^{\alpha\beta}
\end{equation}
represents the standard Maxwell contribution, while the curvature-dependent interaction term is given by
\begin{eqnarray}
T^{(I)}_{\mu\nu} &=& \frac{1}{2} F^{\alpha\beta} F^{\gamma\lambda} g_{\mu\nu} R_{\alpha\gamma\beta\lambda} 
- \frac{3}{2} F^{\beta\gamma} F_{\nu}^{\;\; \alpha} R_{\mu\beta\alpha\gamma} 
- \frac{3}{2} F^{\beta\gamma} F_{\mu}^{\;\; \alpha} R_{\nu\beta\alpha\gamma} \nonumber \\ 
&& -\frac{1}{2} F_{\nu}^{\;\; \alpha} \nabla_\alpha \nabla_\lambda F_{\mu}^{\;\; \lambda} 
-\frac{1}{2} F_{\mu}^{\;\; \alpha} \nabla_\alpha \nabla_\lambda F_{\nu}^{\;\; \lambda} 
- \nabla_\beta F_{\nu}^{\;\; \gamma} \nabla_\gamma F_{\mu}^{\;\; \beta} \nonumber \\ 
&& - \nabla_\beta F_{\mu}^{\;\; \beta} \nabla_\lambda F_{\nu}^{\;\; \lambda} 
- \frac{1}{2} F_{\nu}^{\;\; \alpha} \nabla_\lambda \nabla_\alpha F_{\mu}^{\;\; \lambda} 
- \frac{1}{2} F_{\mu}^{\;\; \alpha} \nabla_\lambda \nabla_\alpha F_{\nu}^{\;\; \lambda}.
\end{eqnarray}

Varying the action~(\ref{action}) with respect to $A_\mu$ yields the modified Maxwell equations,
\begin{equation}\label{EOM-Maxwell}
    \nabla_{\mu} \left( -\frac{1}{2} \alpha F^{\mu\nu} + 2\epsilon F^{\alpha\beta} R_{\alpha}{}^{\mu}{}_{\beta}{}^{\nu} \right) = 0.
\end{equation}

We consider a static, spherically symmetric ansatz for the metric,
\begin{equation}\label{metric}
ds^{2} = -e^{-2H(r)}f(r)dt^{2} + \frac{dr^{2}}{f(r)} + r^{2}(d\theta^{2} + \sin^{2}\theta\, d\phi^{2}),
\end{equation}
where \( H(r) \) is a redshift function.

For the gauge potential we adopt the electrostatic configuration,
\begin{equation}\label{background}
\mathbf{A} = h(r) \, dt.
\end{equation}

Substituting Eqs.~\eqref{metric} and~\eqref{background} into the field equations yields the $tt$-component,
\begin{equation}\label{tt-comp}
	\begin{split}
		&-4 f(r) - 4r\bigl(\Lambda r + f'(r)\bigr) \\
		&+ e^{2H(r)} \kappa \Bigl[ r h'(r) \Bigl( h'(r)\bigl(4f'(r)(\epsilon + 4\epsilon r H'(r)) - r(\alpha + 4\epsilon f''(r))\bigr) + 4\epsilon r f'(r) h''(r) \Bigr) \\
		&+ 8\epsilon f(r) \Bigl( r^2 [h''(r)]^2 + [h'(r)]^2 \bigl(1 + r\bigl( H'(r)(4 + r H'(r)) + 2r H''(r) \bigr)\bigr) \\
		&\quad + r h'(r) \bigl(4(1 + r H'(r)) h''(r) + r h^{(3)}(r)\bigr) \Bigr) \Bigr] = 0.
	\end{split}
\end{equation}

For $\epsilon = 0$, this reduces to the minimally coupled limit,
\begin{equation}\label{ttq20}
4 f_0(r) + 4r\bigl(f_0'(r) + \Lambda r\bigr) + \kappa \alpha r^2 e^{2H_0(r)} [h_0'(r)]^2 = 0.
\end{equation}

The $rr$-component of Einstein's equations reads,
\begin{equation}\label{rr-comp}
	\begin{split}
		&4 f(r) + 4\Lambda r^2 + 4r f'(r) - 8 f(r) r H'(r) \\
		&+ e^{2H(r)} \kappa r h'(r) \Bigl[ -4\epsilon r\bigl(f'(r) - 2f(r) H'(r)\bigr) h''(r) \\
		&+ h'(r) \Bigl( -4f'(r)\bigl(\epsilon + 4\epsilon r H'(r)\bigr) + r\bigl(\alpha + 4\epsilon f''(r)\bigr) \\
		&+ 8\epsilon f(r)\bigl(H'(r) + 2r[H'(r)]^2 - r H''(r)\bigr) \Bigr) \Bigr] = 0,
	\end{split}
\end{equation}

which for $\epsilon = 0$ simplifies to
\begin{equation}
4r\bigl(\Lambda r + f_0'(r)\bigr) + e^{2H_0(r)} \alpha \kappa r^2 [h_0'(r)]^2 + 4f_0(r)\bigl(1 - 2r H_0'(r)\bigr) = 0.
\end{equation}

Subtracting this result from Eq.~(\ref{ttq20}) yields $8 f_0(r) r H_0'(r) = 0$, from which we conclude that $H_0(r)$ must be constant. For simplicity we set $H_0 = 0$.

The $\theta\theta$-component becomes
\begin{align}
	&f'(r)\bigl(4 - 6r H'(r)\bigr) + 2r\bigl(2\Lambda + f''(r)\bigr) \nonumber\\
	&- e^{2H(r)} \kappa r [h'(r)]^2 \Bigl[\alpha + 2\epsilon\bigl(-3f'(r)H'(r) + f''(r) + 2f(r)([H'(r)]^2 - H''(r))\bigr)\Bigr] \nonumber\\
	&+ 4f(r) \Bigl[H'(r)\bigl(-1 + r H'(r)\bigr) - r H''(r)\Bigr] = 0.
\end{align}

Finally, the nontrivial component of the Maxwell equation~(\ref{EOM-Maxwell}) takes the form
\begin{equation}
	\begin{aligned}
		r h''(r) &\Bigl[\alpha + 2\epsilon\Bigl(-3f'(r)H'(r) + f''(r) + 2f(r)\bigl([H'(r)]^2 - H''(r)\bigr)\Bigr)\Bigr] \\
		&+ h'(r)\Bigl[(8\epsilon f(r) - 2\epsilon r f'(r))[H'(r)]^2 + 4\epsilon f(r) r [H'(r)]^3 \\
		&+ H'(r)\Bigl(-12\epsilon f'(r) + r\bigl(\alpha - 4\epsilon f''(r) + 4\epsilon f(r) H''(r)\bigr)\Bigr) \\
		&+ 2\Bigl(\alpha + \epsilon\bigl(2f''(r) + r\bigl(-5f'(r)H''(r) + f^{(3)}(r) - 2f(r)(2H''(r) + r H^{(3)}(r))\bigr)\bigr)\Bigr)\Bigr] = 0.
	\end{aligned}
\end{equation}

The general solution for $h(r)$ can be expressed as
\begin{equation}\label{hhe}
h(r)= C_1 \int^{r} \frac{e^{-H(u)}}{\alpha u^2 + \epsilon B_1(u)}\, du + C_2,
\end{equation}
where
\begin{equation}
B_1(u)= u^2 \Bigl( -6 f'(u) H'(u) + 4 f(u) [H'(u)]^2 + 2 f''(u) - 4 f(u) H''(u) \Bigr).
\end{equation}

Now, we expand the metric and gauge functions as
\begin{align}
f(r) &= f_0(r) + \epsilon f_1(r), \label{f} \\
h(r) &= h_0(r) + \epsilon h_1(r), \label{h} \\
H(r) &= H_0(r) + \epsilon H_1(r). \label{H}
\end{align}

At zeroth order in $\epsilon$, Eq.~\eqref{hhe} yields
\begin{equation}
h_0(r) = C_2 + C_1 \int^r \frac{1}{\alpha u^2}\, du 
        = C_2 - \frac{C_1}{\alpha r},
\end{equation}
where we choose $C_1 = -\alpha Q$ and $C_2 = -Q/r_h$ to satisfy regularity at the black hole horizon $r = r_h$. Consequently,
\begin{equation}
	h_0(r) = Q\left(\frac{1}{r} - \frac{1}{r_h}\right).
\end{equation}

Solving the $\epsilon = 0$ limit of Einstein's equations gives the Reissner--Nordström--AdS-type metric function
\begin{equation}\label{f0}
f_0(r) = 1 - \frac{2m_0}{r} - \frac{\Lambda r^2}{3} + \frac{\kappa \alpha Q^2}{4 r^2},
\end{equation}
with the mass parameter
\begin{equation}\label{m0}
m_0 = \frac{r_h}{2} - \frac{\Lambda r_h^3}{6} + \frac{\kappa \alpha Q^2}{8 r_h}.
\end{equation}

Using relations (\ref{f}), (\ref{h}) and (\ref{H})  then expanding the $tt$-component of Einstein's equations to first order in $\epsilon$ and collecting all terms at order $\epsilon$, gives
\begin{equation}\label{tt2}
	\begin{split}
		&-4 f_1(r) - 4 r f_1'(r) + 8\kappa f_0(r) [h_0'(r)]^2 - 2\alpha\kappa H_1(r) r^2 [h_0'(r)]^2 \\
		&+ 4\kappa r f_0'(r) [h_0'(r)]^2 - 2\alpha\kappa r^2 h_0'(r) h_1'(r) - 4\kappa r^2 [h_0'(r)]^2 f_0''(r) \\
		&+ 32\kappa f_0(r) r h_0'(r) h_0''(r) + 4\kappa r^2 f_0'(r) h_0'(r) h_0''(r) + 8\kappa f_0(r) r^2 [h_0''(r)]^2 \\
		&+ 8\kappa f_0(r) r^2 h_0'(r) h_0'''(r) = 0.
	\end{split}
\end{equation}

Similarly, the $rr$-component expanded to $\mathcal{O}(\epsilon)$ yields
\begin{equation}\label{rr2}
	\begin{aligned}
		4 f_1(r) &+ 4 r f_1'(r) + 2\alpha\kappa H_1(r) r^2 [h_0'(r)]^2 - 4\kappa r f_0'(r) [h_0'(r)]^2 \\
		&+ 2\alpha\kappa r^2 h_0'(r) h_1'(r) - 8 f_0(r) r H_1'(r) + 4\kappa r^2 [h_0'(r)]^2 f_0''(r) \\
		&- 4\kappa r^2 f_0'(r) h_0'(r) h_0''(r) = 0.
	\end{aligned}
\end{equation}

Subtracting Eq.~\eqref{rr2} from Eq.~\eqref{tt2} eliminates $f_1$ and $h_1$, leading to
\begin{equation}\label{Ettq2-Errq2}
	\begin{split}
		8\kappa f_0(r) [h_0'(r)]^2 &- 8 f_0(r) r H_1'(r) + 32\kappa f_0(r) r h_0'(r) h_0''(r) \\
		&+ 8\kappa f_0(r) r^2 [h_0''(r)]^2 + 8\kappa f_0(r) r^2 h_0'(r) h_0'''(r) = 0.
	\end{split}
\end{equation}

Solving Eq.~\eqref{Ettq2-Errq2} for $H_1'(r)$ and integrating gives
\begin{equation}
	H_1(r) = C_3 + \int^{r} \frac{1}{u} \Bigl[ 
	\kappa [h_0'(u)]^2 + 
	4u\kappa h_0'(u) h_0''(u) + 
	u^2\kappa [h_0''(u)]^2 + 
	u^2\kappa h_0'(u) h_0'''(u)
	\Bigr] du,
	\label{eq:H1_solution}
\end{equation}
which, upon inserting $h_0(u) = Q(1/u - 1/r_h)$, simplifies to
\begin{equation}
H_1(r) = C_3 - \frac{3\kappa Q^2}{4 r^4}.
\end{equation}

To preserve asymptotic flatness and a unit speed of light on the AdS boundary~\cite{Mahapatra:2016dae}, we set $C_3 = 0$, giving
\begin{equation}
H_1(r) = -\frac{3\kappa Q^2}{4 r^4}.
\end{equation}

The first-order correction to the gauge field follows from the Maxwell equation and reads
\begin{equation}
\begin{aligned}
h_1(r) &= \alpha Q \int^{r} \frac{\alpha H_1(u) + 2 f_0''(u)}{\alpha^2 u^2} \, du \\
&= -\frac{9\kappa Q^3}{20 r^5} + \frac{2 m_0 Q}{\alpha r^4} + \frac{4\Lambda Q}{3\alpha r} + a,
\end{aligned}
\end{equation}
where $a$ is an integration constant.

Now, Eq.~\eqref{rr2} yields the first-order correction to the metric function,
\begin{equation}\label{f1}
f_1(r) = \frac{b}{r} - \frac{2\alpha\kappa^2 Q^4}{5 r^6} + \frac{7 m_0 \kappa Q^2}{2 r^5} - \frac{2\kappa Q^2}{r^4} + \frac{4\kappa Q^2 \Lambda}{3 r^2},
\end{equation}
where $b$ is an integration constant.

The full metric function up to $\mathcal{O}(\epsilon)$ therefore becomes
\begin{equation}
\begin{aligned}
f(r) &= 1 - \frac{2m_0}{r} - \frac{\Lambda r^2}{3} + \frac{\kappa \alpha Q^2}{4 r^2} \\
&\quad +  \Bigl(\frac{b}{r} - \frac{2\alpha\kappa^2 Q^4}{5 r^6} + \frac{7 m_0 \kappa Q^2}{2 r^5} - \frac{2\kappa Q^2}{r^4} + \frac{4\kappa Q^2 \Lambda}{3 r^2}\Bigr)\epsilon.
\end{aligned}
\end{equation}

To determine the integration constants $a$ and $b$, we introduce a perturbed horizon radius to ensure consistent first-order thermodynamics. We define the shifted horizon as
\begin{equation}
r_h' = r_h + \epsilon r_h^{(1)},
\end{equation}
where $r_h$ is the unperturbed horizon radius and $\epsilon r_h^{(1)}$ is its first-order correction.

Applying the condition $h(r_h') = 0$ and expanding in $\epsilon$ gives
\begin{align}
h(r_h + \epsilon r_h^{(1)}) 
&= h_0(r_h) + \epsilon\bigl[ h_0'(r_h) r_h^{(1)} + h_1(r_h) \bigr] + \mathcal{O}(\epsilon^2) = 0,
\end{align}
which yields the horizon shift
\begin{equation}
r_h^{(1)} = -\frac{h_1(r_h)}{h_0'(r_h)}.
\end{equation}
Thus the perturbed horizon to $\mathcal{O}(\epsilon)$ reads
\begin{equation}\label{ph}
r_h' = r_h - \epsilon \frac{h_1(r_h)}{h_0'(r_h)}.
\end{equation}

Imposing $h(r_h') = 0$ to first order leads to
\begin{equation}
a = -\frac{4\Lambda Q}{3\alpha r_h} + \frac{9\kappa Q^3}{20 r_h^5} - \frac{2 m_0 Q}{\alpha r_h^4}.
\end{equation}
Interestingly, we note that the condition $h_1(r_h) = 0$ holds automatically; therefore, the constant $a$ can be determined without invoking the shifted horizon. However, this may not hold at higher orders in the perturbative expansion.

For internal consistency of the perturbation scheme, the horizon shift derived from $h(r)$ must match that obtained from the metric function $f(r)$. This matching imposes the condition
\begin{equation}\label{con}
\frac{h_1(r_h)}{h_0'(r_h)} = \frac{f_1(r_h)}{f_0'(r_h)}.
\end{equation}

Since $h_1(r_h) = 0$, Eq.~(\ref{con}) implies $f_1(r_h) = 0$, which fixes the constant $b$ as
\begin{equation}
b = -\frac{4\kappa\Lambda Q^2}{3 r_h} + \frac{2\kappa^2\alpha Q^4}{5 r_h^5} + \frac{2\kappa Q^2}{r_h^3} - \frac{7\kappa Q^2 m_0}{2 r_h^4}.
\end{equation}

Substituting $b$ and using the expression for $m_0$ from Eq.~(\ref{m0}), we obtain the final form of the gauge potential and metric function to first order in $\epsilon$:

\begin{equation}
\begin{aligned}
h(r) &= \frac{Q}{r} - \frac{Q}{r_h} \\
&\quad +  \Bigg[ \frac{4\Lambda Q}{3\alpha r} - \frac{\Lambda Q}{\alpha r_h} - \frac{r_h^3\Lambda Q}{3\alpha r^4} - \frac{9\kappa Q^3}{20 r^5} + \frac{\kappa Q^3}{4 r^4 r_h} \\
&\qquad\quad + \frac{\kappa Q^3}{5 r_h^5} + \frac{r_h Q}{\alpha r^4} - \frac{Q}{\alpha r_h^3} \Bigg]\epsilon,
\end{aligned}
\end{equation}

\begin{equation}
\begin{aligned}
f(r) &= 1 - \frac{2M}{r} - \frac{\Lambda r^2}{3} + \frac{\frac{1}{4}\alpha\kappa Q^2 + \frac{4}{3}\epsilon\kappa\Lambda Q^2}{r^2} - \frac{2\kappa\epsilon Q^2}{r^4} \\
&\quad + \frac{1}{r^5}\Bigg( -\frac{7r_h^3\kappa\Lambda Q^2}{12} + \frac{7r_h\kappa Q^2}{4} + \frac{7\kappa^2\alpha Q^4}{16 r_h} \Bigg)\epsilon \\
&\quad - \frac{2\alpha\kappa^2 Q^4}{5 r^6}\epsilon,
\end{aligned}
\end{equation}

where the ADM mass is given by
\begin{equation}\label{ADM}
M = \frac{r_h}{2} - \frac{\Lambda r_h^3}{6} + \frac{\kappa\alpha Q^2}{8 r_h}
      + \Bigg( \frac{3\kappa\Lambda Q^2}{8 r_h} - \frac{\kappa Q^2}{8 r_h^3} + \frac{3\kappa^2\alpha Q^4}{160 r_h^5} \Bigg)\epsilon.
\end{equation}

Expanding the time–time component of the metric, \( g_{tt} = e^{-2H(r)} f(r) \), perturbatively in the coupling parameter \( \epsilon \) and comparing the result with the standard asymptotic expansion of an asymptotically AdS spacetime,
\begin{equation}
g_{tt} = 1 - \frac{\Lambda r^2}{3} - \frac{2M}{r} + \cdots ,
\end{equation}
confirms that the ADM mass of the black hole, to first order in \( \epsilon \), is indeed given by Eq.~(\ref{ADM}) \cite{Arnowitt:1962hi, Henneaux:1985tv}.

In general, the coefficient of the \(1/r\) term in \(g_{tt}\) does not always directly determine the physical mass. This is particularly relevant in scenarios involving scalar fields with slow falloff, higher-curvature terms, or modified boundary conditions, where the asymptotic structure receives additional contributions \cite{Henneaux:2006hk, Skenderis:2002wp}. In such cases, holographic renormalization provides the most systematic and reliable method for extracting conserved quantities \cite{Balasubramanian:1999re, deHaro:2000vlm}. However, in our perturbative setup, the corrections at order \(\epsilon\) only shift the value of \(M=m_0 + m_1 \epsilon \) without altering the asymptotic falloff structure. Consequently, the mass information is still encoded entirely in the coefficient of the \(1/r\) term\cite{Sadeghi:2025ae128a}.

%--------------------------------------------------------------------------

\section{Conventional Thermodynamic Analysis}
\label{sec3}

We begin by investigating the thermodynamics of the system using conventional methods, establishing essential results that will be used in the subsequent topological analysis. In the extended phase space, the enthalpy of the system is identified with the ADM mass derived earlier, expressed in terms of thermodynamic pressure via the relation $\Lambda = -8\pi P$. From this point onward, we take $\kappa=1$ and $\alpha=1$ in all calculations. To avoid confusion with the redshift function $H(r)$, we denote the enthalpy by $\mathcal{H}$, which reads
\begin{equation}
\mathcal{H} = \frac{r_h}{2} + \frac{4 r_h^{3} \pi P}{3} + \frac{\kappa \alpha Q^{2}}{8 r_h}
             + \left(-\frac{3 \kappa Q^{2} \pi P}{r_h} + \frac{3 \kappa^{2} Q^{4} \alpha}{160 r_h^{5}} - \frac{\kappa Q^{2}}{8 r_h^{3}}\right) \epsilon .
\end{equation}

The Hawking temperature is obtained from the surface gravity formula,
\begin{equation}\label{Temp}
\begin{split}
T &= \frac{1}{2\pi} \left[ \frac{1}{\sqrt{g_{rr}}} \frac{d}{dr} \sqrt{-g_{tt}} \right] \Bigg|_{r = r_h} 
   = \frac{e^{-H(r_h)} f'(r_h)}{4 \pi} \\
  &= 2 P r_{h} - \frac{\kappa \alpha Q^{2}}{16 \pi r_{h}^{3}} + \frac{1}{4 \pi r_{h}}
    + \left(-\frac{\kappa Q^{2} P}{2 r_{h}^{3}} + \frac{\kappa^{2} Q^{4} \alpha}{64 \pi r_{h}^{7}} - \frac{\kappa Q^{2}}{16 \pi r_{h}^{5}}\right) \epsilon \\
  &= \left( \frac{ \partial \mathcal{H}/\partial r_h }{ \partial S/\partial r_h } \right)_{P,Q},
\end{split}
\end{equation}
The entropy of the black hole, computed via the thermodynamic identity
\begin{equation}
S = \int \frac{1}{T} \frac{\partial \mathcal{H}}{\partial r_h} \, dr_h 
    = \pi r_h^2 - \frac{\kappa \pi Q^2 \epsilon}{r_h^2},
\end{equation}
shows a correction at first order in $\epsilon$ relative to the usual area law.

Using Eq.~\eqref{Temp}, the pressure can be expressed in terms of the temperature and the horizon radius as
\begin{equation}
P = -\frac{64 \pi T r_{h}^{7} - Q^{4} \alpha \epsilon \kappa^{2} + 4 Q^{2} \alpha \kappa r_{h}^{4} + 4 Q^{2} \epsilon \kappa r_{h}^{2} - 16 r_{h}^{6}}
        {32 \pi r_{h}^{4} \left(\kappa Q^{2} \epsilon - 4 r_{h}^{4}\right)}.
\end{equation}
The denominator in these expressions has potential zeros that could indicate singularities. Importantly, the critical points (where phase transitions occur) and the physically admissible parameter regime are all located away from these singularities, guaranteeing that thermodynamic quantities remain well-behaved in the region of interest.

Finally, the Helmholtz free energy, obtained from the Legendre transform $F = \mathcal{H} - T S$, reads to first order in $\epsilon$
\begin{equation}\label{GR}
F = \frac{r_{h}}{4} - \frac{2 r_{h}^{3} \pi P}{3} + \frac{3 \kappa \alpha Q^{2}}{16 r_{h}}
    + \left(-\frac{\kappa Q^{2} \pi P}{2 r_{h}} - \frac{19 \kappa^{2} Q^{4} \alpha}{320 r_{h}^{5}} + \frac{3 \kappa Q^{2}}{16 r_{h}^{3}}\right) \epsilon .
\end{equation}

Another important quantity that characterizes the local stability of the system is the heat capacity at constant pressure,
\begin{equation}
C_P =\left[\frac{\partial \mathcal{H}}{\partial r_h} \left(\frac{\partial T}{\partial r_h}\right)^{-1}\right]_{P}= \frac{ \left( \dfrac{3\kappa Q^{2} \pi P}{r_{h}^{2}} 
               - \dfrac{3\kappa^{2} Q^{4} \alpha}{32 r_{h}^{6}} 
               + \dfrac{3\kappa Q^{2}}{8 r_{h}^{4}} \right) \epsilon 
               + 4\pi P r_{h}^{2} - \dfrac{\kappa \alpha Q^{2}}{8 r_{h}^{2}} + \dfrac12 }
             { 2P + \dfrac{3\kappa \alpha Q^{2}}{16\pi r_{h}^{4}} 
               - \dfrac{1}{4\pi r_{h}^{2}}
               + \left( \dfrac{3\kappa Q^{2} P}{2 r_{h}^{4}} 
               - \dfrac{7\kappa^{2} Q^{4} \alpha}{64\pi r_{h}^{8}} 
               + \dfrac{5\kappa Q^{2}}{16\pi r_{h}^{6}} \right) \epsilon } .
\end{equation}

In the canonical ensemble, our analysis reveals that the system satisfies the conditions
\begin{equation}
    \left.\frac{\partial P}{\partial r_h}\right|_{T = T_c, \, r_h = r_c} = 0, \qquad 
    \left.\frac{\partial^2 P}{\partial r_h^2}\right|_{T = T_c, \, r_h = r_c} = 0,
\end{equation}
which confirm a van der Waals-like phase structure. Obtaining closed-form analytical expressions for the critical pressure $P_c$, temperature $T_c$, and horizon radius $r_c$ proved intractable; therefore, we determined these quantities numerically.

\begin{figure}[H]
\centering
\subfloat[Pressure vs. horizon radius at fixed $T = T_c$ for different charges.]{\includegraphics[width=7cm]{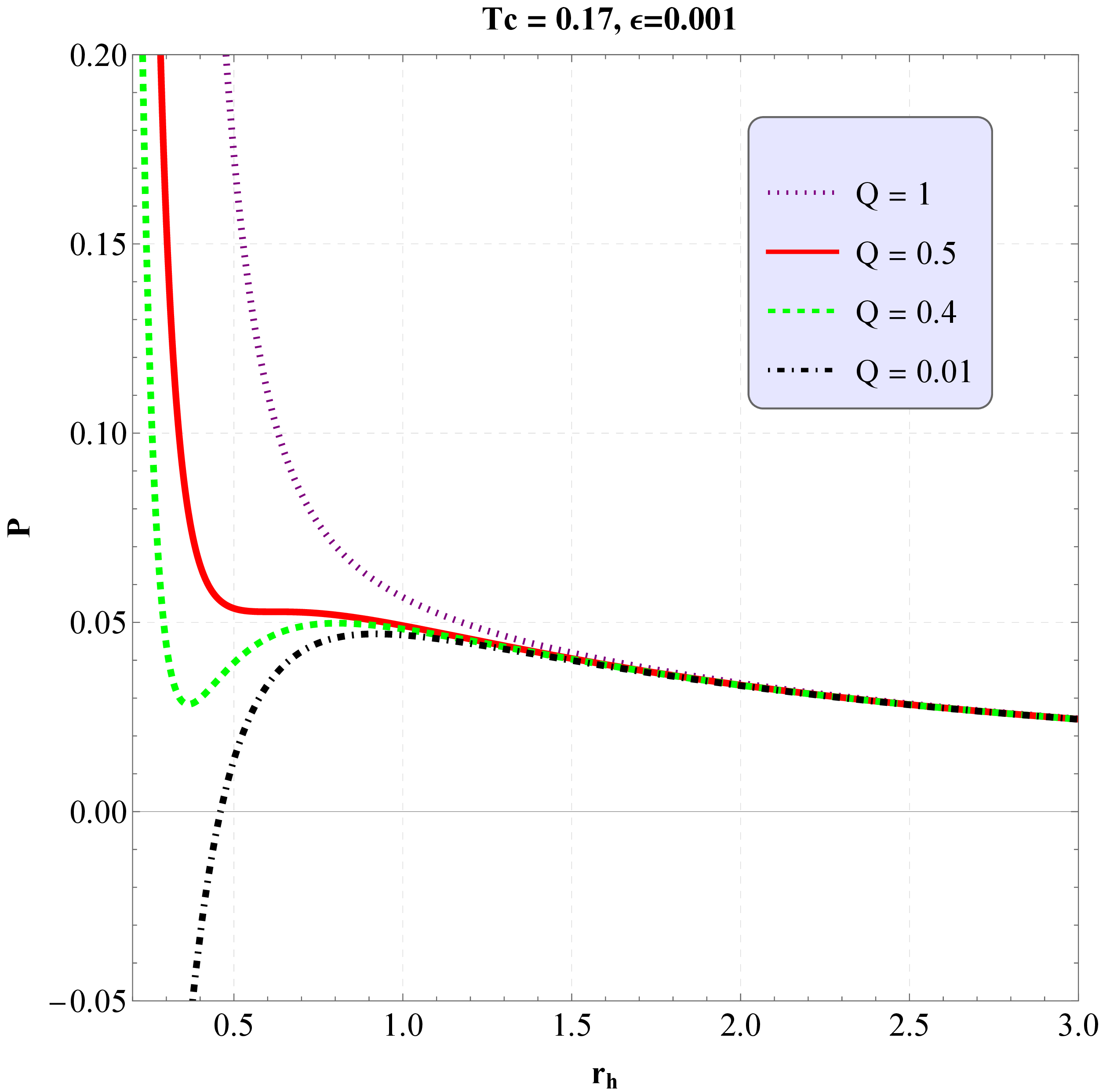}}\quad
\subfloat[Temperature vs. horizon radius at fixed $P = P_c$ for different charges.]{\includegraphics[width=7cm]{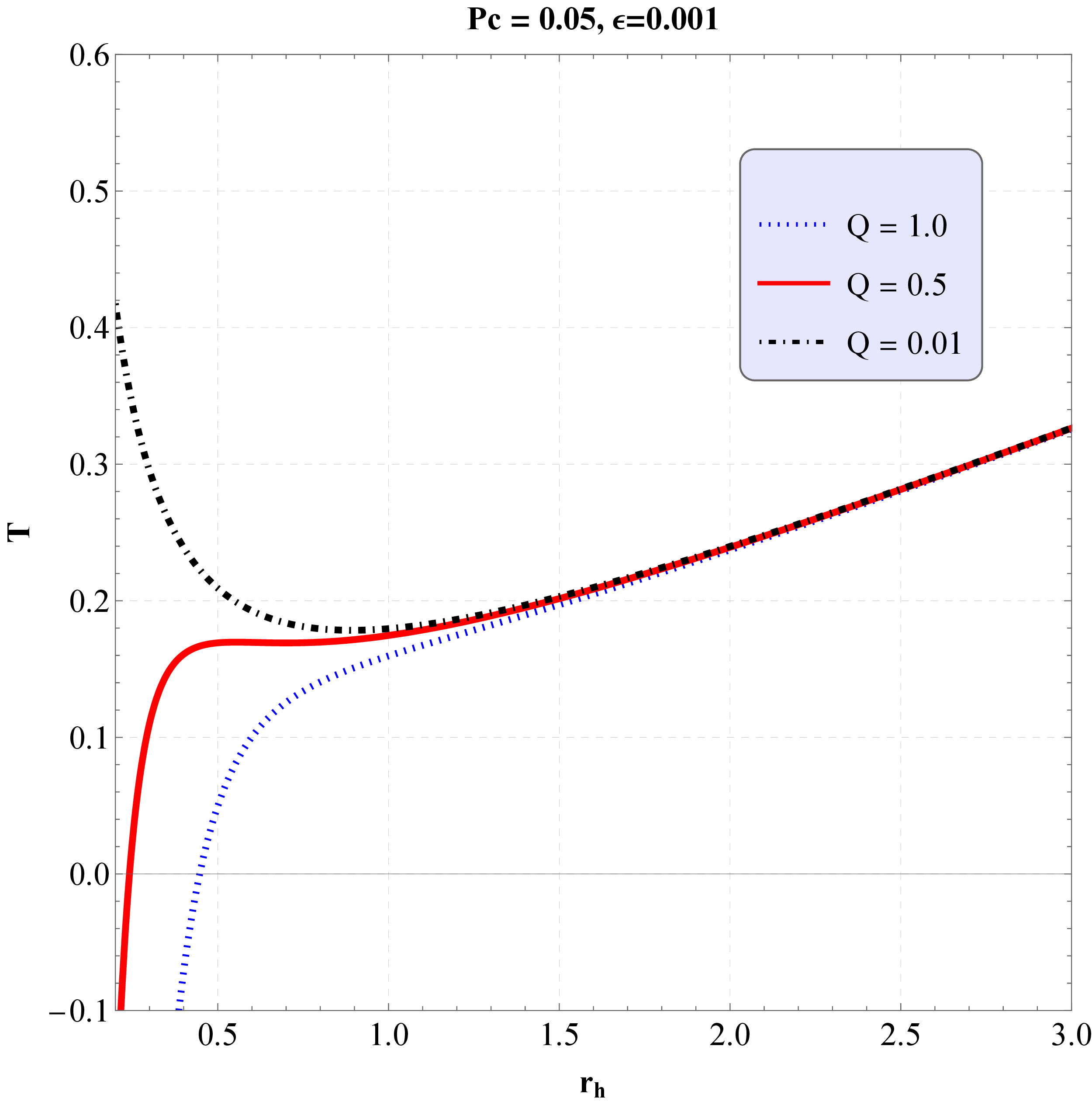}}
\caption{Van der Waals-like behavior in the $P$–$r_h$ and $T$–$r_h$ diagrams. 
         (a) Pressure as a function of $r_h$ at the critical temperature $T_c = 0.17$ for several charge values.
         (b) Temperature as a function of $r_h$ at the critical pressure $P_c = 0.05$ for several charge values.\label{fig:pt}}
\end{figure}

Fixing $Q = 0.5$ and $\epsilon = 0.001$, we find the critical values
\begin{equation}
r_c \approx 0.61, \qquad T_c \approx 0.17, \qquad P_c \approx 0.05.
\end{equation}
The resulting critical behavior is illustrated in Fig.~\ref{fig:pt} for varying values of the Maxwell charge $Q$.

\begin{figure}[H]
\centering
\subfloat[Heat capacity at constant pressure for $P=0.02 < P_c$, showing three distinct phases. For $P = 0.08$, only one stable phase exists.]{\includegraphics[width=7cm]{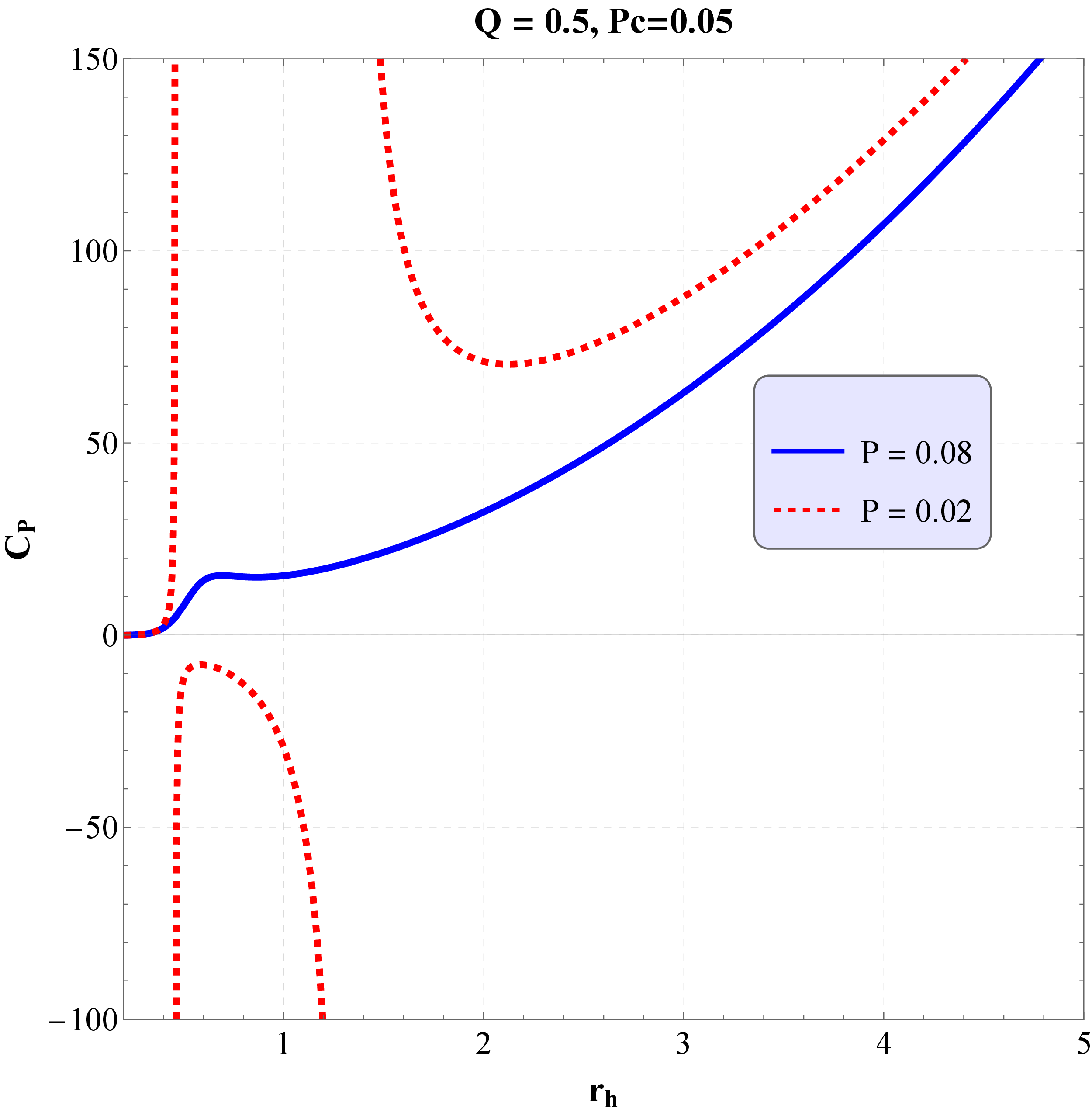}}\quad
\subfloat[Helmholtz free energy exhibiting the swallow-tail structure typical of a van der Waals fluid.]{\includegraphics[width=7cm]{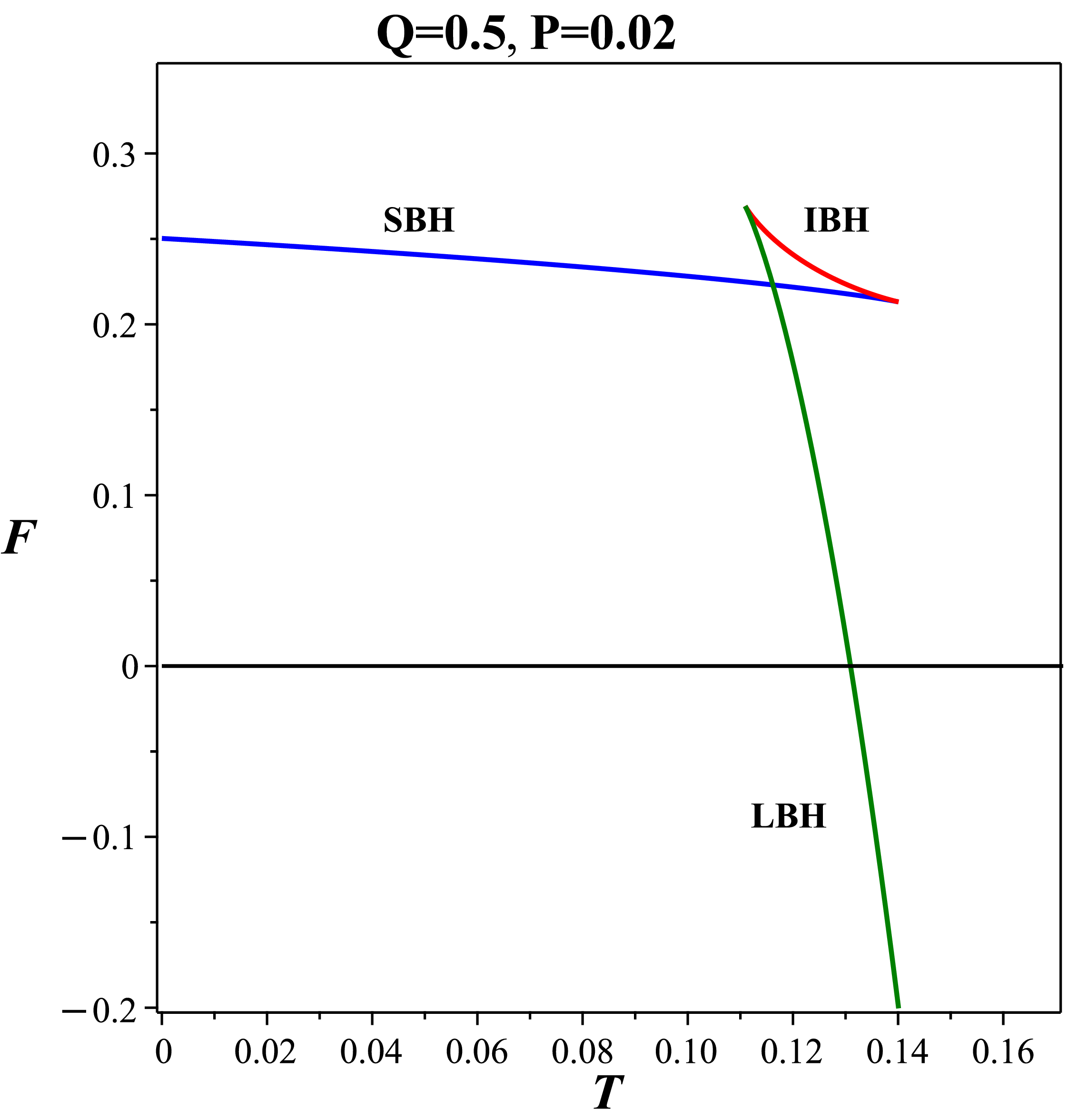}}
\caption{Local and global thermodynamic behavior of the system. 
         (a) Heat capacity for subcritical pressure, revealing stable/unstable/stable phase structure.
         (b) Helmholtz free energy displaying the swallow-tail shape characteristic of van der Waals-like phase transitions.\label{fig:c1gt1}}
\end{figure}

The left panel of Fig.~\ref{fig:c1gt1} displays the black hole heat capacity at constant thermodynamic pressure, \(C_P\), which serves as a diagnostic for local thermodynamic stability within the extended phase space. For pressures below the critical value, the diagram reveals three distinct branches: a stable small black hole branch with positive \(C_P\), a stable large black hole branch with positive \(C_P\), and an unstable intermediate black hole branch where \(C_P\) is negative. This structure is the hallmark of a first-order small/large black hole phase transition, analogous to the liquid-gas transition. Above the critical pressure (the blue curve), the distinction between small and large black hole phases disappears, leaving a single, thermodynamically stable black hole branch for all horizons. In this supercritical regime, \(C_P\) remains positive and continuous, indicating no phase transitions and a smooth evolution of the black hole's thermodynamic response as its temperature changes. The van der Waals-like behavior is thus suppressed, yielding a globally stable ``black hole fluid.'' At the critical pressure, the unstable intermediate branch vanishes, and the stable small and large black hole branches merge at a single critical point. Here, the phase transition becomes second-order, and \(C_P\) diverges with a characteristic critical exponent rather than exhibiting the discontinuous jump seen subcritically.

The right panel presents the global behavior of the Helmholtz free energy for fixed charge \(Q = 0.5\) and pressure \(P = 0.02\) (a subcritical value). The emergence of the characteristic swallow-tail structure confirms the van der Waals-like first-order phase transition in the system. This feature arises from the intersection of branches representing the small, unstable intermediate, and large black hole phases. As temperature increases, the system undergoes a transition toward the thermodynamically preferred large black hole phase, which minimizes the free energy and ultimately dominates the ensemble at high temperatures.

The phase structure has so far been examined using conventional thermodynamic methods; these results will serve as input for the topological analysis that follows. A more comprehensive investigation of the standard thermodynamic behavior could be pursued, but the focus here remains on the topological approach. In the next section, we will demonstrate how the topological method, through its off-shell free energy formulation, confirms this phase structure.

\section{Topological Formalism for Phase Transitions}
\label{sec4}
The topological analysis of black hole thermodynamics provides a geometric perspective that classifies phase transitions through global properties of the thermodynamic parameter space. This approach, pioneered by Wei, Liu, and Mann \cite{Wei:2022dzw}, treats black hole solutions as topological defects characterized by winding numbers, offering a universal classification scheme independent of specific black hole metrics. The method builds upon Duan's $\phi$-mapping topological current theory \cite{Duan:1998kw,Duan:1984} and the off-shell free energy formalism introduced by York \cite{York:1986it} in the Euclidean path integral approach to quantum gravity. Recent applications have revealed four distinct topological classes for black hole thermodynamics \cite{Wei:2022dzw}.  

Following the Euclidean path integral approach of Gibbons--Hawking--York \cite{York:1986it,Gibbons:1976ue,Hawking:1976de}, a black hole in a cavity with fixed boundary temperature $1/\tau$ is examined. The generalized off-shell free energy is defined as:
\begin{equation}
    \mathcal{F} = M - \frac{S}{\tau},
    \label{eq:offshell}
\end{equation}
where $M$ denotes the black hole mass, $S$ the entropy, and $\tau$ the inverse temperature of the cavity. Within the framework of the extended phase space, the black hole mass is interpreted as enthalpy and, following the convention adopted here, is represented by $\mathcal{H}$.
 Only when $\tau = \beta \equiv 1/T$, with $T$ being the Hawking temperature, does $\mathcal{F}$ become an on-shell quantity corresponding to equilibrium states.

To apply topological analysis, an auxiliary angular coordinate $\Theta \in (0, \pi)$ is introduced, and a two-component vector field $\phi = (\phi^{r_h}, \phi^\Theta)$ is defined as:
\begin{align}
    \phi^{r_h} &= \frac{\partial \mathcal{F}}{\partial r_h} = \frac{\partial S}{\partial r_h}\left(\frac{1}{\beta} - \frac{1}{\tau}\right), \\
    \phi^\Theta &= \cot\Theta \csc\Theta,
    \label{vcomp}
\end{align}
where $r_h$ is the horizon radius~\cite{Wei:2022dzw}. The zero points of $\phi$ correspond to black hole equilibrium states, which satisfy $\partial_{r_h}\mathcal{F} = 0$, or equivalently, $\beta = \tau$ with $\Theta = \pi/2$.

Using Duan's $\phi$-mapping topological current theory \cite{Duan:1998kw,Duan:1984}, the topological current is defined as:
\begin{equation}
    j^\mu = \frac{1}{2\pi}\epsilon^{\mu\nu}\epsilon_{ab}\partial_\nu n^a \partial_\rho n^b, \quad \mu,\nu,\rho = 0,1,
    \label{eq:topocurrent}
\end{equation}
where $n^a = \phi^a/||\phi||$ with $a=1,2$ is the normalized vector field. This current is conserved, $\partial_\mu j^\mu = 0$, and can be expressed in the form
\begin{equation}
    j^\mu = \delta^2(\phi)J^\mu\left(\frac{\phi}{x}\right),
\end{equation}
indicating that it is non-vanishing only at the zero points of $\phi$.

Here, $J^\mu\left(\frac{\phi}{x}\right)$ represents the Jacobian vector or topological current density. This structure makes $j^\mu$ a topological current whose integral over a closed surface counts the winding number (or topological charge) of the zero points~\cite{Wei:2022dzw2}. At each zero point $z_i$, we compute the winding number (Brouwer degree):
\begin{equation}\label{wi}
w_i = \int j^0 \, d^2 x =  \beta_i \eta_i
\end{equation}
where 
\begin{equation}
j^0 =  \beta_i \, \eta_i \; \delta^2(\vec{x} - \vec{z}_i).
\end{equation}

The positive integer $\beta_i$ (the Hopf index) counts the number of loops traced by $\phi^a$ in its internal space when $x^\mu$ encircles the zero point $z_i$ of the field i.e $\phi(\vec{x})\vert_{\vec{x}=\vec{z}_i}=0$. The Brouwer degree $\eta_i = \mathrm{sign}\left(J^0(\phi/x)_{z_i}\right) = \pm 1$ indicates the orientation of the vector field at the zero.

The global topological invariant for a black hole system is the sum of all winding numbers:
\begin{equation}
    W = \sum_{i=1}^{N} w_i,
    \label{eq:totalwinding}
\end{equation}
where $N$ is the total number of isolated zero points within the region of interest.

 The winding number characterizes the local topological property:
\begin{itemize}
    \item $w_i = +1$: Stable black hole state (positive heat capacity)
    \item $w_i = -1$: Unstable black hole state (negative heat capacity)
   \end{itemize}

To calculate the winding number associated with a given zero point of the vector field \(\phi\), the deflection of the vector field along a contour \(C_i\) can also be used~\cite{WeiLiu2020}. A closed contour \(C_i\) is constructed to enclose the zero point in the \((r_h, \Theta)\)-plane. A convenient parametric form for such a contour is given by:
\begin{equation}
\label{eq:contour_param}
\begin{aligned}
r_h(\vartheta) &= a\cos\vartheta + z_0, \\
\Theta(\vartheta) &= b\sin\vartheta + \frac{\pi}{2},
\end{aligned}
\quad 0 \leq \vartheta \leq 2\pi,
\end{equation}
where \(a\) and \(b\) are adjustable parameters chosen so that the contour encircles the desired zero point \(z_0\). 
The deflection of the normalized vector field \(n = \phi/\|\phi\|\) along the contour is quantified by:
\begin{equation}
\label{ome1}
\Omega(\vartheta) = \int_0^\vartheta \epsilon_{ab} \, n^a \partial_{\vartheta'} n^b \, d\vartheta',
\end{equation}
where \(\epsilon_{ab}\) is the two-dimensional Levi-Civita symbol~\cite{Ahmed:2022kyv}. The total topological charge \(\Upsilon\) enclosed by the contour \(C_i\) is then obtained from the total deflection after one full circuit:
\begin{equation}
\label{ome2}
\Upsilon = \frac{\Omega(2\pi)}{2\pi}.
\end{equation}
If the contour \(C_i\) encloses a single zero point \(z_i\), the quantity \(\Upsilon\) equals the winding number \(w_i\). If \(C_i\) instead encloses all zero points in the parameter space, \(\Upsilon\) recovers the global topological number \(W = \sum_i w_i\). This method provides a direct, computable link between the local topological structure of the vector field and the global classification of black hole solutions.

\begin{table}[htbp]
\centering
\caption{Asymptotic behaviors of the inverse temperature \(\beta(r_h)\).}
\label{tab:asymptotic}
\begin{tabular}{|c|c|c|}
\hline
\textbf{Case} & \(\beta(r_h \to r_m)\) & \(\beta(r_h \to \infty)\) \\
\hline
I & 0 & \(\infty\) \\
II & \(\infty\) & \(\infty\) \\
III & \(\infty\) & 0 \\
IV & 0 & 0 \\
\hline
\end{tabular}
\end{table}

The topological classification of black hole thermodynamics is determined by the asymptotic behavior of the inverse temperature $\beta(r_h)$ at the boundaries of the parameter space \cite{Wei:2022dzw}. The relevant limits are $r_h \to r_m$ (small black hole) and $r_h \to \infty$ (large black hole), where $r_m$ is the minimal horizon radius. As shown in Table~\ref{tab:asymptotic}, this leads to four distinct cases for $\beta$. The boundaries of the region of interest are defined as:
\begin{align}
I_1 &= \{r_h = \infty,\; \Theta \in (0, \pi)\}, \nonumber \\
I_2 &= \{r_h \in (\infty, r_m),\; \Theta = \pi\}, \nonumber \\
I_3 &= \{r_h = r_m,\; \Theta \in (\pi, 0)\}, \nonumber \\
I_4 &= \{r_h \in (r_m, \infty),\; \Theta = 0\}. \nonumber
\end{align}

The sign of the vector field component $\phi^{r_h}$ at these boundaries follows directly from its definition in (\ref{vcomp}). Given the fixed cavity temperature $\tau$ and the fact that entropy increases with horizon radius ($\partial S/\partial r_h > 0$), the asymptotic direction is:
\begin{itemize}
    \item $\phi^{r_h} \rightarrow$ (rightward) when $\beta \to 0$,
    \item $\phi^{r_h} \leftarrow$ (leftward) when $\beta \to \infty$.
\end{itemize}
This behavior, combined with the fixed vertical directions of $\phi^\Theta$ on the other boundaries, determines the total topological number $W$ obtained by integrating along a closed contour enclosing the entire parameter space.

Consequently, there are four universal topological classes, denoted $W^{XY}$. Here, $X$ represents the total topological number $W$ (where $1 \equiv +1$), and $Y = \pm$ indicates the winding number of the innermost black hole branch. Their defining properties are summarized in Table~\ref{tab:topclasses}.

\begin{table}[ht]
\centering
\caption{Topological classification of black hole thermodynamics. The notation \(W^{XY}\) encodes the total topological number \(X\) and the winding sign \(Y\) of the innermost black hole branch. The patterns \([w_{\mathrm{in}}, w_{\mathrm{out}}]\) denote the winding numbers of the innermost and outermost branches, respectively.}
\label{tab:topclasses}
\begin{tabular}{|c|c|c|c|c|}
\hline
\textbf{Class} & \textbf{Notation} & \(\mathbf{[w_{\mathrm{in}}, w_{\mathrm{out}}]}\) & \textbf{Small BH ($r_h \to r_m$)} & \textbf{Large BH ($r_h \to \infty$)} \\
\hline
Class I & $W^{1-}$ & $[-,-]$ & Unstable & Unstable \\
Class II & $W^{0+}$ & $[+,-]$ & Stable & Unstable \\
Class III & $W^{1+}$ & $[+,+]$ & Stable & Stable \\
Class IV & $W^{0-}$ & $[-,+]$ & Unstable & Stable \\
\hline
\end{tabular}
\end{table}

Each class corresponds to a canonical black hole solution and a specific thermodynamic interpretation:
\begin{itemize}
    \item \textbf{Class I ($W^{1-}$)}: Represented by the Schwarzschild black hole. It contains at least one unstable state; any additional states appear in pairs with alternating stability.
    \item \textbf{Class II ($W^{0+}$)}: Exemplified by the Reissner-Nordström (RN) black hole. It is characterized by a stable small black hole and an unstable large black hole.
    \item \textbf{Class III ($W^{1+}$)}: Represented by the RN-AdS black hole. It contains at least one stable state; any additional states appear in pairs with alternating stability.
    \item \textbf{Class IV ($W^{0-}$)}: Exemplified by the Schwarzschild-AdS (SAdS) black hole. It is characterized by an unstable small black hole and a stable large black hole, akin to the Hawking–Page transition.
\end{itemize}
Classes II and IV both have a total topological number $W=0$ but are distinguished by the stability of their innermost branch ($+$ vs. $-$), leading to their distinct notation and thermodynamic behavior.

Points where $\partial\beta/\partial r_h = 0$ correspond to \textit{degenerate zero points} of the vector field $\phi$, where multiple zero points coincide. At these points, the net winding number vanishes, signaling a topological transition.

This condition is directly linked to thermodynamic phase transitions. Starting from the definition $\beta = 1/T$, we have:
\begin{equation}
\frac{\partial\beta}{\partial r_h} = \frac{\partial}{\partial r_h}\left(\frac{1}{T}\right) = -\frac{1}{T^2}\frac{\partial T}{\partial r_h}.
\end{equation}

Then, we can write:
\[
\frac{\partial T}{\partial r_h} = \frac{\partial T}{\partial S}\,\frac{\partial S}{\partial r_h}.
\]

Combining this with the definition of the heat capacity $C = T\,(\partial S/\partial T) = T\,(\partial T/\partial S)^{-1}$, we obtain:

\begin{equation}
\frac{\partial\beta}{\partial r_h} = -\frac{1}{T^2}\frac{\partial S}{\partial r_h}\frac{\partial T}{\partial S} 
= -\frac{1}{T}\frac{\partial S}{\partial r_h}\frac{1}{C}.
\label{eq:bet}
\end{equation}

Equation~(\ref{eq:bet}) shows that $\partial\beta/\partial r_h = 0$ occurs when:

\begin{enumerate}
    \item $C \to \infty$: The heat capacity diverges. This defines a \textit{Davies point}, which marks a thermodynamic instability and a change in local stability.
    \item Alternatively, $\partial S/\partial r_h = 0$, but this is unphysical for standard black holes where entropy grows with $r_h$.
\end{enumerate}

Thus, degenerate zero points coincide with Davies points in the thermodynamic ensembles considered here. At these points, the thermodynamic stability changes and the system undergoes a phase transition.

In the topological framework, degenerate points represent \textit{bifurcation points} on the defect curve $\beta(r_h)$. Based on the behavior of black hole states as the cavity temperature $\tau$ varies, they are classified into two types:

\begin{itemize}
    \item \textbf{Generation Points}: At a generation point, a pair of black hole states (one stable, $w = +1$, and one unstable, $w = -1$) emerges as $\tau$ increases. This occurs when the defect curve $\beta(r_h)$ develops a local minimum that dips below a critical temperature. For example, in Reissner–Nordström (RN) black holes, as the cavity temperature rises above a critical value, two states, a small stable black hole and a large unstable one, are generated from a single extremal configuration.
    
    \item \textbf{Annihilation Points}: At an annihilation point, a pair of black hole states merges and disappears as $\tau$ varies. This occurs when a local maximum in $\beta(r_h)$ rises above a critical temperature. In Schwarzschild–AdS (SAdS) black holes, for instance, as the temperature decreases, the small unstable and large stable black holes converge and annihilate at a critical temperature.
\end{itemize}

The two types are distinguished by the curvature of $\beta(r_h)$ at the degenerate point:
\begin{itemize}
    \item Generation points: $\displaystyle \frac{\partial^2\beta}{\partial r_h^2} > 0$ (local minimum)
    \item Annihilation points: $\displaystyle \frac{\partial^2\beta}{\partial r_h^2} < 0$ (local maximum)
\end{itemize}

Degenerate points play a crucial role in the topological classification:
\begin{itemize}
    \item In classes $W^{0+}$ and $W^{0-}$ (which have total winding number $W = 0$), degenerate points are necessary because these classes contain pairs of black hole states with opposite winding numbers.
    \item In classes $W^{1-}$ and $W^{1+}$ (with $W = \pm 1$), degenerate points may also appear in pairs (one generation and one annihilation) if intermediate black hole states exist.
\end{itemize}

In the following subsection, we apply this topological formalism to our four-dimensional charged AdS black hole with $F^{\alpha\beta}F^{\gamma\lambda}R_{\alpha\gamma\beta\lambda}$ coupling. 

\subsection{Topological Phase Structure with $F^{\alpha\beta}F^{\gamma\lambda}R_{\alpha\gamma\beta\lambda}$ Coupling }

Following the topological formalism, we first construct the off-shell free energy for the system:

\begin{equation}\label{free}
\mathcal{F} = \mathcal{H} - \frac{S}{\tau} = \frac{r_{h}}{2} + \frac{4 r_{h}^{3} \pi P}{3} + \frac{\kappa \alpha Q^{2}}{8 r_{h}} + \left( -\frac{3 \kappa Q^{2} \pi P}{r_{h}} + \frac{3 \kappa^{2} Q^{4} \alpha}{160 r_{h}^{5}} - \frac{\kappa Q^{2}}{8 r_{h}^{3}} \right) \epsilon - \frac{\pi r_{h}^{2} - \frac{ Q^{2} \kappa \pi}{r_{h}^{2}}\epsilon}{\tau}.
\end{equation}

The  vector field component $\phi^{r_h}=\frac{\partial \mathcal{F}}{\partial r_h}$ is given by

\begin{equation}\label{phi1}
\phi^{r_h} = \frac{\partial \mathcal{F}}{\partial r_h} = \frac{1}{2} + 4 r_{h}^{2} \pi P - \frac{\kappa \alpha Q^{2}}{8 r_{h}^{2}} + \left( \frac{3 \kappa Q^{2} \pi P}{r_{h}^{2}} - \frac{3 \kappa^{2} Q^{4} \alpha}{32 r_{h}^{6}} + \frac{3 \kappa Q^{2}}{8 r_{h}^{4}} \right) \epsilon - \frac{2 \pi r_{h} + \frac{2  Q^{2} \kappa \pi}{r_{h}^{3}}\epsilon}{\tau},
\end{equation}

and \(\phi^\Theta = -\cot\Theta\csc\Theta\). 

For the parameter values \(Q = 0.5\) and \(\epsilon = 0.001\), we obtain the critical values \(P_c = 0.05\) and \(T_c = 0.17\). Consequently, we have \(\tau_c = 1/T_c \approx 5.88\). For pressures below the critical value \(P_c = 0.05\) and for \(\tau > \tau_c\), the system exhibits three topological defects. At \(\tau = 8\), these correspond to three equilibrium phases located at \(\mathrm{ZP}_1 = 0.35\), \(\mathrm{ZP}_2 = 0.73\), and \(\mathrm{ZP}_3 = 2.25\).

The winding number associated with each zero point, for non-degenerate points, is given by
\[
w_i = \eta_i = \mathrm{sign}\!\left(\frac{\partial^2 \mathcal{F}}{\partial r_h^2}\right)\!\Bigg|_{r_h = \mathrm{ZP}_i}.
\]
This simplification arises because the divergence of the topological current produces a delta function localized at the defects, weighted by this Jacobian. Consequently, a positive second derivative corresponds to a winding number \(\eta_i = +1\) (indicating a vortex-like defect), whereas a negative second derivative yields \(\eta_i = -1\) (an anti-vortex defect). The non-degeneracy condition ensures that these defects are isolated and that their topological charges are well-defined and robust.

The three defects carry winding numbers \(w_1 = +1\), \(w_2 = -1\), and \(w_3 = +1\), respectively. Their sum yields the global topological invariant
\[
W = \sum_{i=1}^3 w_i = +1,
\]
which, according to the classification scheme in Table~\ref{tab:topclasses}, places the system in Class III (denoted \(W^{1+}\)). This class is characterized by at least one stable black hole state and, when multiple states exist, they appear in pairs with alternating stability. In our case, the sequence of winding numbers from small to large horizon radii is \([+,-,+]\), indicating that the innermost and outermost black holes are stable (\(w = +1\)), while the intermediate one is unstable (\(w = -1\)).

The asymptotic behavior of the inverse temperature \(\beta(r_h)\) further confirms this classification. Analysis of our system shows that \(\beta \to \infty\) as \(r_h \to r_m\) (the minimal radius) and \(\beta \to 0\) as \(r_h \to \infty\), corresponding to Case~III in Table~\ref{tab:asymptotic}. 

For clarity, the detailed asymptotic limits are as follows. At large horizon radius, \(r_h \to \infty\), the dominant term gives
\[
\beta \simeq \frac{1}{2 P r_h} \to 0^+,
\]
so that \(\beta \to 0\) at the AdS boundary, as expected.

On the other hand, from the thermodynamic behavior one can observe that there is a small minimum horizon radius for which $T=0$ and, consequently, $\beta$ diverges. This is exactly what we expect in class~III. According to Ref.~\cite{Wei:2022dzw}, this asymptotic pattern—together with the total winding number \(W = +1\), uniquely identifies the system as belonging to Class~III (\(W^{1+}\)), which contains RN-AdS-type black holes.

\begin{figure}[H]
\centering
\subfloat[Unit vector field for \(P = 0.02 < P_c\), \(Q = 0.5\), and \(\tau = 8\), exhibiting three topological defects. The boundary behavior identifies this configuration as belonging to Class III.]{\includegraphics[width=7.5cm]{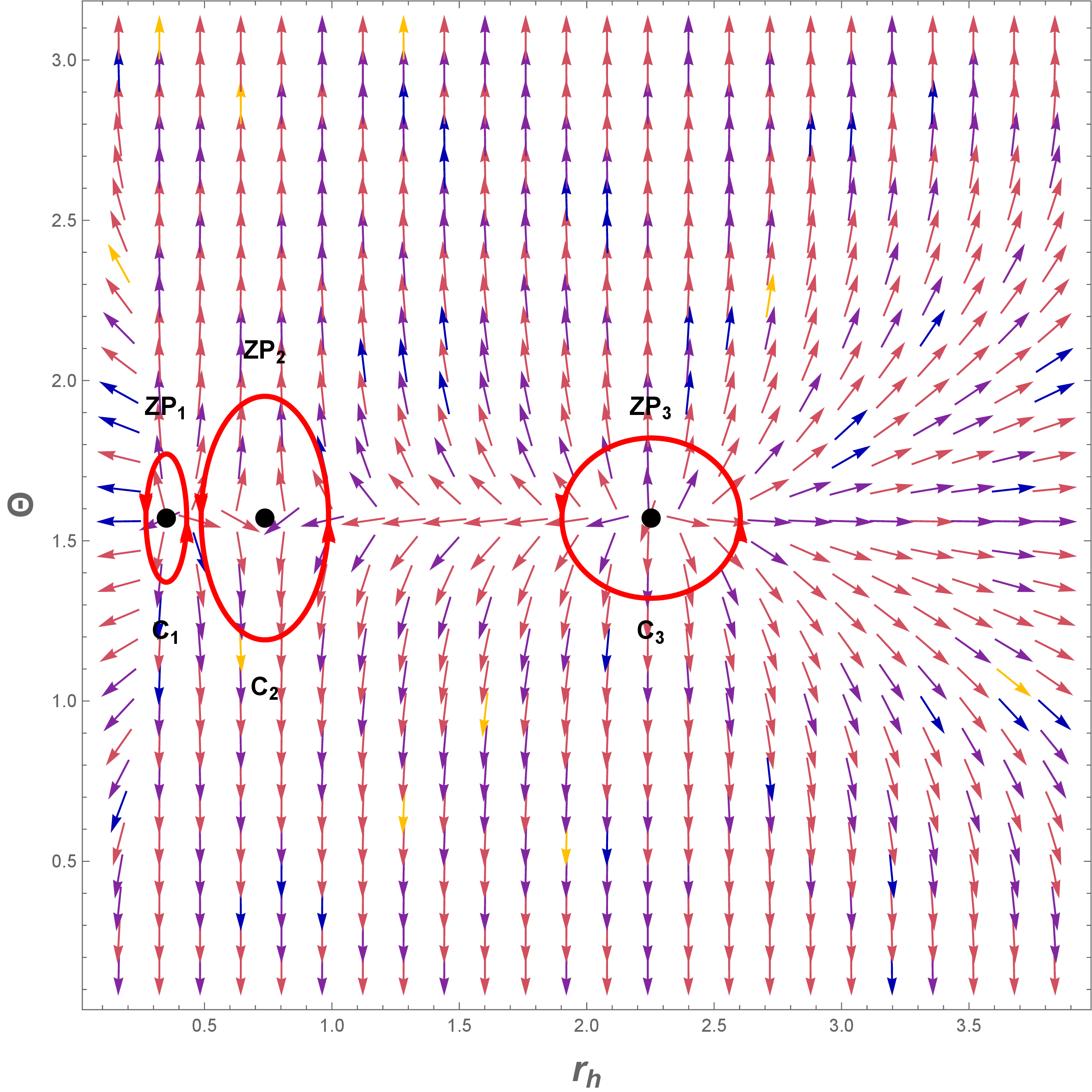}}\quad
\subfloat[\(r_h\)–\( \tau\) diagram for \(P = 0.02\) and \(Q = 0.5\), showing the generation point \(a\) and annihilation point \(b\). The total topological number remains invariant within each colored region.]{\includegraphics[width=7.2cm]{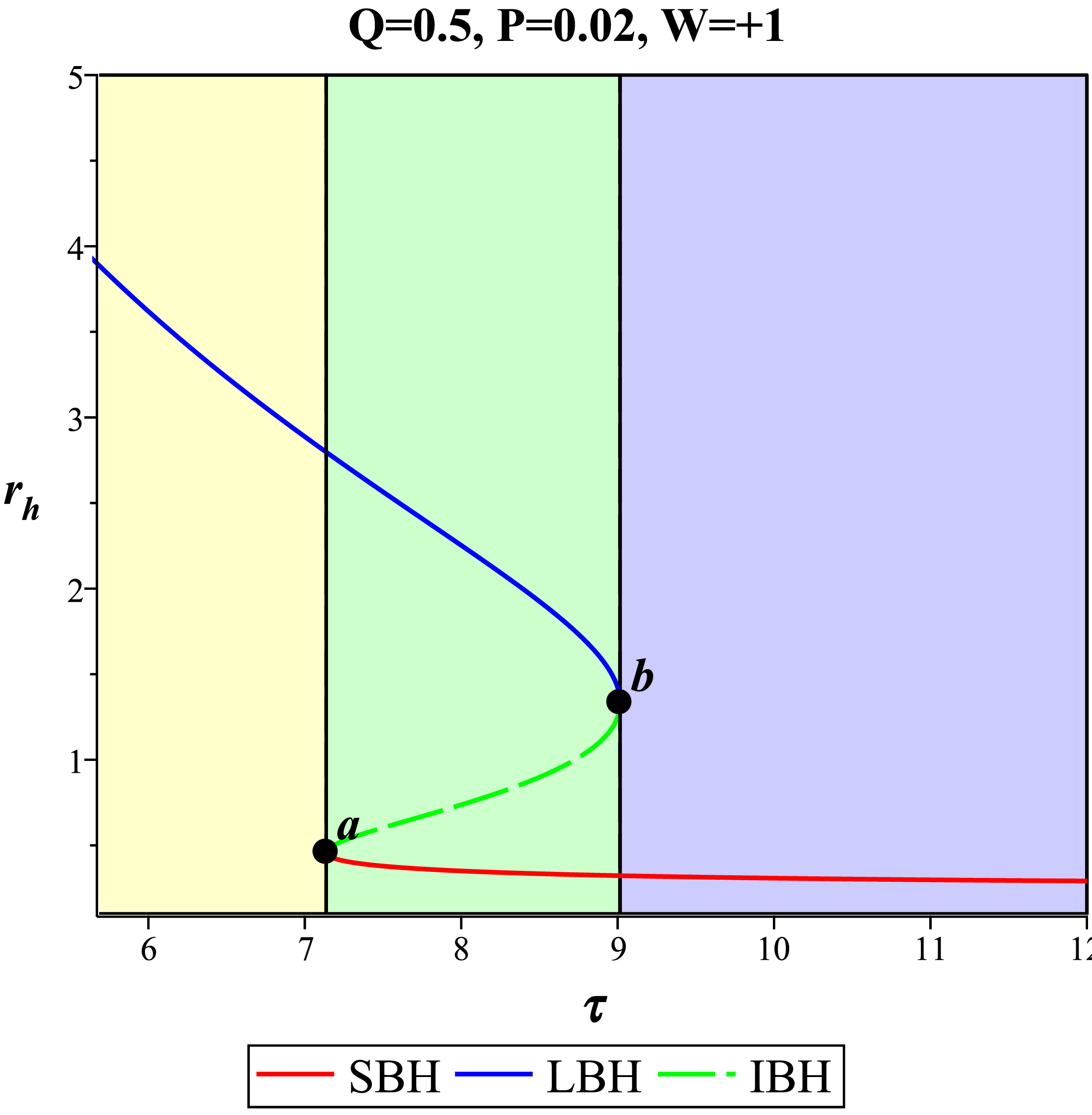}}
\caption{Topological structure and phase diagram for the black hole system. \label{fig:v1tau1}}
\end{figure}

The left panel of Fig.~\ref{fig:v1tau1} illustrates the unit vector field \(n = \phi/\|\phi\|\) in the \((r_h, \Theta)\)-plane and its zeros. The direction of the vector field at the boundaries \(r_h \to r_m\) and \(r_h \to \infty\) is consistent with the asymptotic limits of \(\beta(r_h)\).

Setting \(\phi^{r_h}=0\) yields:
\begin{equation}\label{taurh}
\tau=\frac{64 \pi  (\epsilon  Q^{2} \kappa +r_{h}^{4}) r_{h}^{3}}{96 P \pi  Q^{2} \epsilon  \kappa r_{h}^{4}+128 P \pi  r_{h}^{8}-3 Q^{4} \alpha  \epsilon  \kappa^{2}-4 Q^{2} \alpha  \kappa r_{h}^{4}+12 Q^{2} \epsilon  \kappa r_{h}^{2}+16 r_{h}^{6}}.
\end{equation}
This relation allows us to plot the \(r_h\)--\(\tau\) curve, which reveals the thermodynamic behavior of the system in different parameter regimes.

The right panel displays the \(r_h\)-\(\tau\) curve for \(Q = 0.5\) and \(P = 0.02\), highlighting the generation (point $a$) and annihilation points (point $b$) where \(\partial\beta/\partial r_h = 0\). These degenerate points coincide with Davies points, where the heat capacity diverges, signaling first-order phase transitions between the stable and unstable branches.

The yellow region (for \(\tau < \tau_a\)) corresponds to only a large black hole phase. The intermediate green region consists of three coexisting phases: small, intermediate, and large black holes. The purple region (for \(\tau > \tau_b\)) contains only a small black hole phase. In each region, the total topological number \(W\) remains invariant and equal to \(+1\).

As previously noted, the winding numbers can also be computed independently using the deflection-angle method, which serves as an additional check on the values obtained from the Jacobian. For each zero point \(z_i\), we consider a closed contour \(C_i\) in the \((r_h, \Theta)\) plane that encloses the defect. The total change in the angle of the vector field \(\phi\) along \(C_i\) is then calculated using relations (\ref{ome1}) and (\ref{ome2}).

Furthermore, consider the changes in the components of the vector field \(\phi\) as it varies along each contour \(C_i\) in the \((r_h, \Theta)\) plane. Mapping these variations onto the \((\phi^{r_h}, \phi^\Theta)\) plane yields a closed curve \(\Phi_i\), whose origin corresponds to a zero point of \(\phi\). The sense of rotation, clockwise or counterclockwise of \(\Phi_i\) as the corresponding contour in the \((r_h, \Theta)\) plane is traversed determines the sign of the winding number of the enclosed defect. This geometric construction provides a direct visual confirmation of the topological charges and serves as a consistency check against the Jacobian method.

\begin{figure}[H]
\centering
\subfloat[Mapping of contours to the \(\phi\)-plane]{\includegraphics[width=0.45\textwidth]{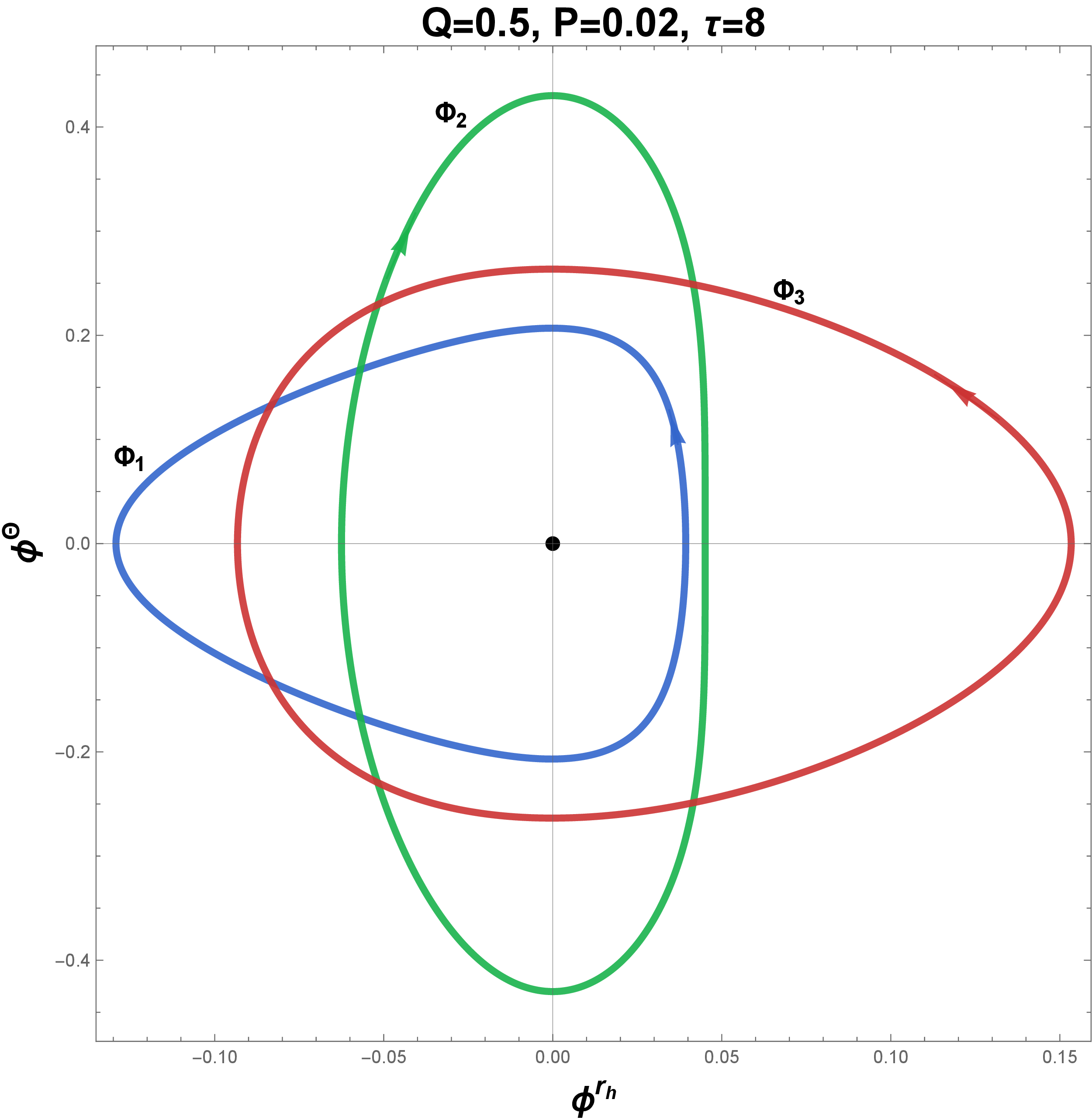}}\hfill
\subfloat[Deflection angle \(\Omega(\vartheta)\) for each contour]{\includegraphics[width=0.45\textwidth]{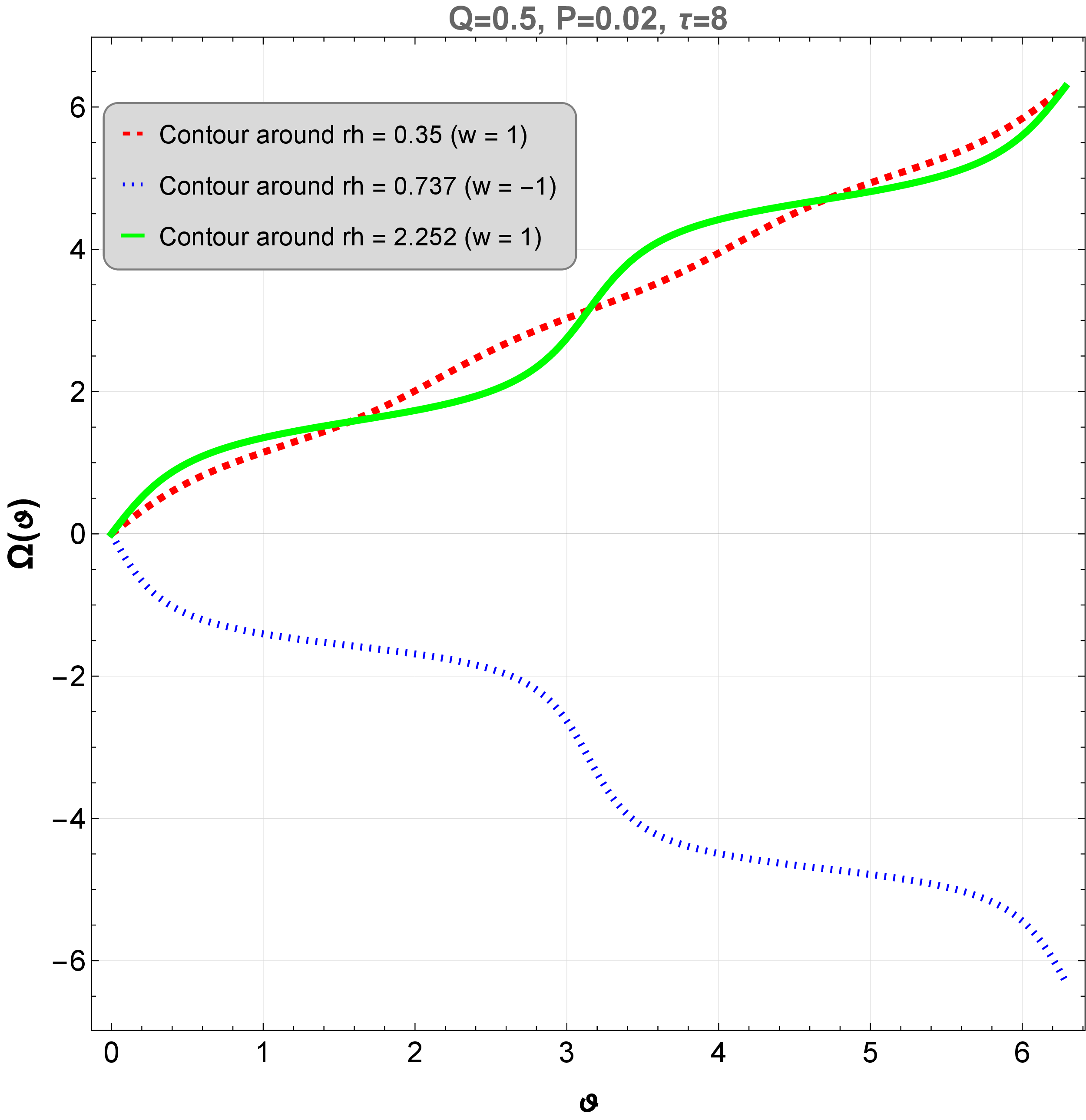}}
\caption{Contour mapping and deflection angle analysis. (a) The curves show the evolution of the \(\phi\)-field components in the \(\phi\)-plane as the contours \(C_i\) in the \((r_h, \Theta)\) plane are traversed. The three zero points in the original coordinate space map to the origin \((0,0)\) in this representation, with the direction of each curve indicating the sign of the corresponding winding number. (b) The evolution of the deflection angle \(\Omega(\vartheta)\) around each zero point provides independent confirmation of the winding numbers.}
\label{fig:con1omega1}
\end{figure}

For the specific parameter choice \(Q = 0.5\) and \(P = 0.02\), this procedure is illustrated in Fig.~\ref{fig:con1omega1}. The left panel shows the contours \(\Phi_i\), which clearly reflect winding numbers of \(+1\), \(-1\), and \(+1\) for the three zero points. The right panel shows the behavior of the deflection angle around each zero point. As expected, for the small and large black holes, \(\Omega(2\pi) = 2\pi\), whereas for the intermediate unstable black hole, \(\Omega(2\pi) = -2\pi\).

For further verification, we examine the vector field structure and the corresponding deflection angle for other values of the Maxwell charge and system pressure. These plots serve to confirm the consistency of the topological assignments across different parameter regimes.

\begin{figure}[H]
\centering
\subfloat[Unit vector field and its zero points for \(Q = 0.8\), \(P = 0.01<P_c\), and \(\tau = 12\).]{\includegraphics[width=0.45\textwidth]{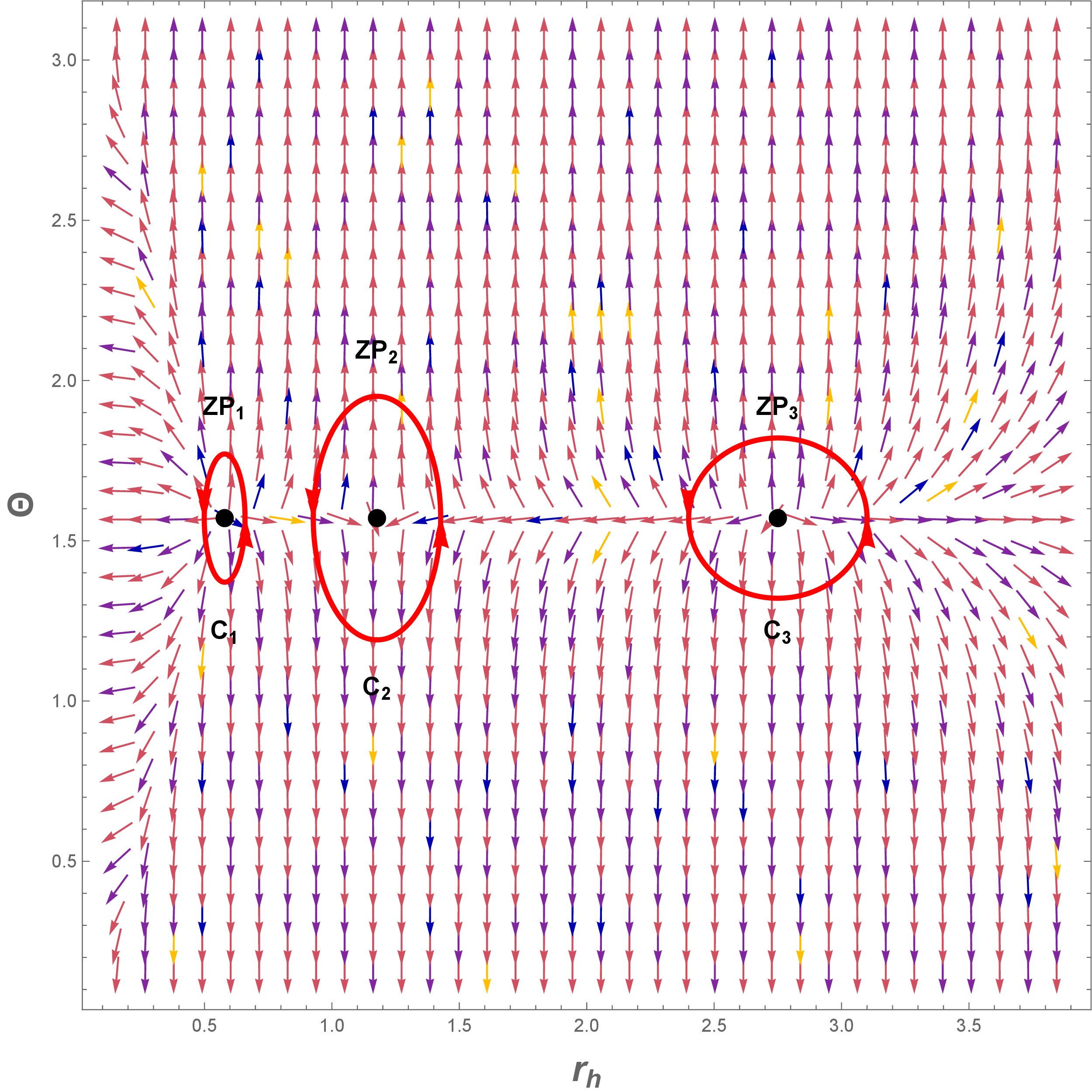}}\hfill
\subfloat[Deflection angle \(\Omega(\vartheta)\) diagrams for contours \(C_1\), \(C_2\), and \(C_3\).]{\includegraphics[width=0.45\textwidth]{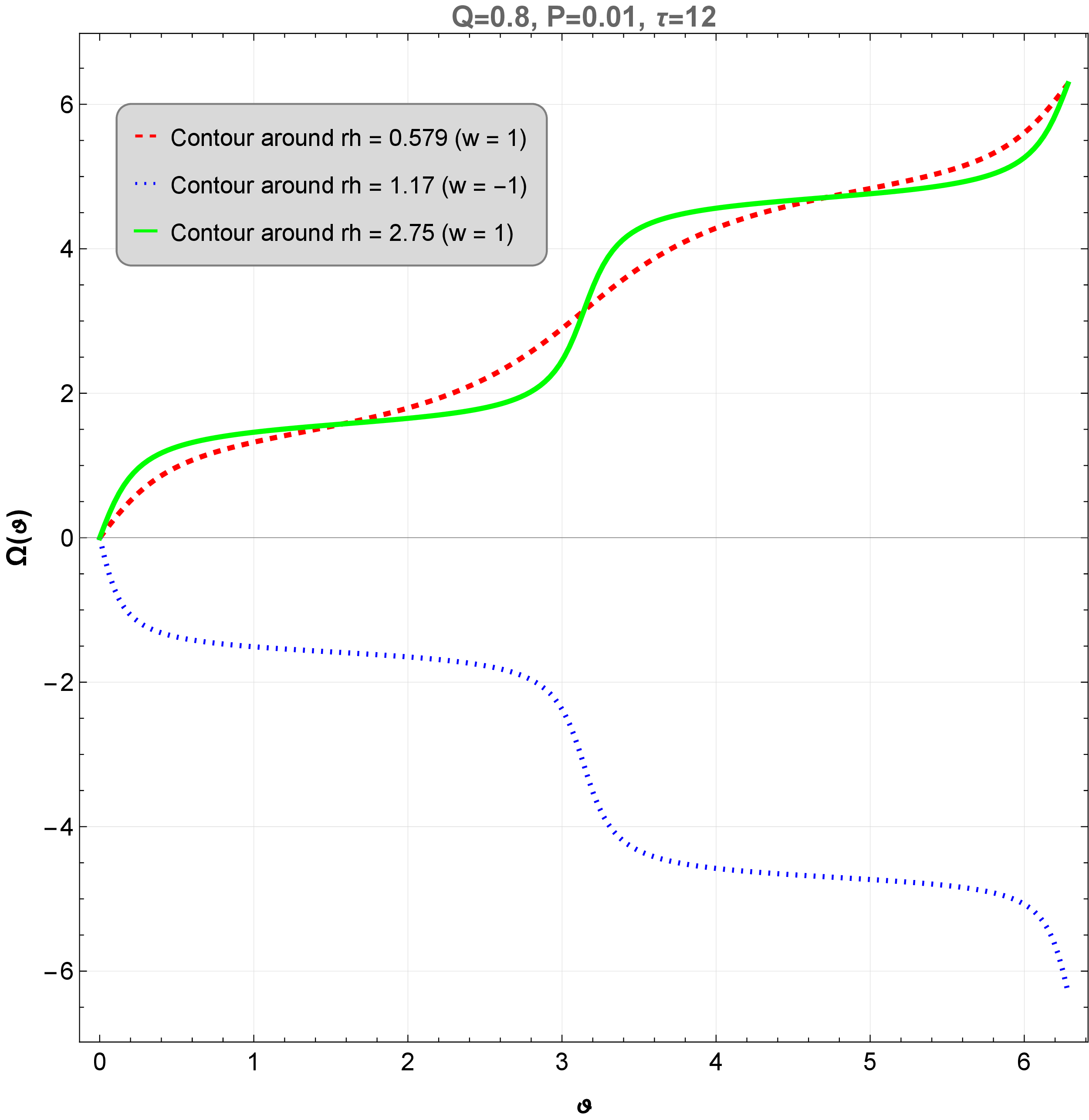}}
\caption{Unit vector field topology and deflection angle analysis. (a) The unit vector field and its boundary behavior for \(Q = 0.8\), \(P = 0.01 < P_c = 0.02\), and \(\tau = 12\) demonstrate that the system belongs to Case III of the topological classification. (b) The evolution of the deflection angle \(\Omega(\vartheta)\) around each zero point independently verifies the winding numbers \((+1, -1, +1)\) associated with the three black hole states.}
\label{fig:v2omega2}
\end{figure}

For \(Q = 0.8\) and \(\epsilon = 0.001\), the critical horizon radius is \(r_c = 0.980\), while the critical pressure and temperature are \(P_c = 0.020\) and \(T_c = 0.0108\), respectively. Therefore, for a pressure value \(P = 0.01\) (below the critical pressure), we expect to observe three zero points (or black hole states) at
\[
\mathrm{ZP}_1 = 0.579,\quad \mathrm{ZP}_2 = 1.177,\quad \mathrm{ZP}_3 = 2.750.
\]
The boundary behavior of the vector field in the left and right panels of Fig.~\ref{fig:v2omega2} clearly indicates that the system belongs to Case III of the topological classification, with winding numbers \(w_1 = +1\), \(w_2 = -1\), and \(w_3 = +1\).

This comprehensive investigation confirms the results of conventional thermodynamics regarding the nature of the phase transitions in this model. The topological approach, based on the off-shell free energy and the associated vector field \(\phi\), provides a robust geometric framework that not only reproduces known thermodynamic classifications but also offers deeper insights into the global structure of the phase space.

%--------------------------------------------------------------------------
\section{Conclusions}
\label{sec5}

\noindent
In this work, we systematically investigate the thermodynamic and topological properties of a four-dimensional charged AdS black hole with a non-minimal gauge-curvature coupling of the form \(F^{\alpha\beta}F^{\gamma\lambda}R_{\alpha\gamma\beta\lambda}\). Using a perturbative approach for the small coupling parameter \(\epsilon\), we derive a consistent black hole solution to first order in \(\epsilon\) and analyze its thermodynamics via both conventional equilibrium methods and the recently developed topological classification formalism.

Our conventional thermodynamic analysis confirms that the system exhibits van der Waals-like critical behavior, characterized by a swallow-tail structure in the free energy and distinct stable/unstable/stable phases below a critical pressure. The critical point \((P_c, T_c, r_c)\) is determined numerically, and local stability is verified through the heat capacity \(C_P\). These results establish the system as a genuine gravitational analogue of a liquid--gas phase transition, reinforcing the robustness of such critical phenomena in extended gravitational theories.

More significantly, we apply the topological classification scheme introduced by Wei, Liu, and Mann \cite{Wei:2022dzw} to our black hole system. By constructing the off-shell free energy and the associated vector field \(\phi\), we identify three zero points, thermodynamic defects, corresponding to the three equilibrium phases. The calculated winding numbers yield the pattern \(w = (+1, -1, +1)\), summing to a global topological invariant \(W = +1\). Together with the asymptotic behaviors \(\beta(r_m) \to \infty\) and \(\beta(\infty) \to 0\), this unequivocally places our system in class III (\(W^{1+}\)) of the universal topological classification. This class, typified by RN-AdS black holes, is characterized by stable innermost and outermost black hole branches separated by an unstable intermediate branch, a structure fully consistent with our conventional thermodynamic results. The fact that a non-minimally coupled AdS black hole falls into the same topological class as RN-AdS black holes suggests a deeper structural universality among charged AdS black holes with stable large-radius phases, regardless of the precise form of gauge-curvature interactions.

Looking forward, several interesting extensions present themselves. It would be worthwhile to explore higher-order perturbative corrections in \(\epsilon\) or to consider the strong-coupling regime where non-perturbative effects may alter the topological class. Additionally, including rotation, higher dimensions, or more general non-minimal couplings could reveal new topological classes or phase transitions not captured in the current setup.

In summary, we have shown that a four-dimensional charged AdS black hole with \(F^{\alpha\beta}F^{\gamma\lambda}R_{\alpha\gamma\beta\lambda}\) coupling exhibits rich thermodynamic criticality and belongs to the universal topological class \(W^{1+}\). Our results reinforce the value of topological tools in gravitational thermodynamics and open new avenues for exploring the geometric nature of black hole phase transitions in modified gravity theories.

%%%%%%%%%%%%%%%%%%%%%%%%%%%5

\vspace{1cm}
\noindent \textbf{Data Availability Statement:} No data were generated or analyzed in this study; therefore, data sharing is not applicable.

%--------------------------------------------------------------------------

\end{document}